# Dark matter halo properties versus local density and cosmic web location

Tze Goh,[1,2]★ Joel Primack,[2]★ Christoph T. Lee,[2] Miguel Aragon-Calvo,[3]
Doug Hellinger,[2] Peter Behroozi[,4] Aldo Rodriguez-Puebla,[3] Elliot Eckholm[2] and
Kathryn Johnston[1]

[1]*Department of Astronomy, Columbia University, 550 West 120th Street, New York, NY 10027, USA*
[2]*Physics Department, University of California, Santa Cruz, 1156 High Street, Santa Cruz, CA 95064, USA*
[3]*Universidad Nacional Autonoma de Mexico, Cto. Exterior S/N, Cd. Universitaria, 04510 Ciudad de Mexico, CDMX, Mexico*
[4]*Department of Astronomy and Steward Observatory, University of Arizona, 933 N Cherry Ave, Tucson, AZ 85721, USA*

## ABSTRACT

We study the effects of the local environmental density and the cosmic web environment (filaments, walls, and voids) on key properties of dark matter haloes using the Bolshoi–Planck $\Lambda$ cold dark matter cosmological simulation. The $z = 0$ simulation is analysed into filaments, walls, and voids using the SpineWeb method and also the VIDE package of tools, both of which use the watershed transform. The key halo properties that we study are the specific mass accretion rate, spin parameter, concentration, prolateness, scale factor of the last major merger, and scale factor when the halo had half of its $z = 0$ mass. For all these properties, we find that there is no discernible difference between the halo properties in filaments, walls, or voids when compared at the same environmental density. As a result, we conclude that environmental density is the core attribute that affects these properties. This conclusion is in line with recent findings that properties of galaxies in redshift surveys are independent of their cosmic web environment at the same environmental density at $z \sim 0$. We also find that the local web environment around galaxies of Milky Way's and Andromeda's masses that are near the centre of a cosmic wall does not appear to have any effect on the properties of those galaxies' dark matter haloes except on their orientation, although we find that it is rather rare to have such massive haloes near the centre of a relatively small cosmic wall.

## 1 INTRODUCTION

The basic structure of the cosmic web was described in the early 1970s as arising from the one-dimensional gravitational collapse of adiabatic fluctuations into pancakes/sheets and subsequently two- and three-dimensional collapse into filaments and nodes/knots (Zel'dovich 1970; Doroshkevich, Shandarin & Zeldovich 1983). These ideas could be realized in detail (e.g. Bond, Kofman & Pogosyan 1996) when the cold dark matter (CDM) theory was developed (Blumenthal et al. 1984) and the $\Lambda$CDM density spectrum of adiabatic fluctuations was supported by the anisotropy structure of the cosmic background radiation and other observational evidence. The modern consensus is now that about 26 per cent of the cosmic density is CDM, $\sim$5 per cent is ordinary (baryonic) matter, and $\sim$69 per cent is dark energy perhaps in the form of a cosmological constant (e.g. Planck Collaboration XIII 2016). The history and evolution of these concepts has recently been summarized in a major conference (van de Weygaert et al. 2016). The various modern methods for determining the structure of the cosmic web have been compared in Libeskind et al. (2018), and reference therein. These methods generally agree on the range of cosmic densities corresponding to voids, with greater dispersion between the different methods in the densities assigned to sheets, filaments, and nodes. The cosmic densities assigned to these various cosmic web locations overlap somewhat, so that haloes at a given environmental density can be in different web environments.

This paper looks at effects of the cosmic web environment on the properties of distinct dark matter haloes (i.e. haloes that are not sub-haloes) in the Bolshoi–Planck cosmological simulation (Klypin et al. 2016; Rodríguez-Puebla et al. 2016) at redshift $z = 0$. The main tool that we use to define the web in the simulations is the Spine of the Web (SpineWeb) (Aragón-Calvo et al. 2010; Aragon-Calvo & Szalay 2013), which starts by identifying voids. The boundaries of voids are walls/sheets, and the boundaries of the sheets are filaments. Other popular methods include the T-web and V-web, in which the cosmic web structures are identified by analysing the tidal and velocity shear fields; for example, voids in the V-web are char-acterized by diverging velocities, while nodes/knots are locations of converging velocities. The methods such as SpineWeb – which do not identify nodes but do identify filaments, walls, and voids – agree that more than 80 per cent of the halo mass is in filaments, with less in walls, and least in voids (Libeskind et al. 2018). The volume fraction assigned to voids is about 40 per cent in methods including SpineWeb and T-web, although the V-web assigns a volume fraction of about 70 per cent to voids (Libeskind et al. 2018). These volume fractions depend heavily on the definition used to classify the environments. In particular, the eigenvalue threshold on the T- or V-web can shift the quantities significantly. To verify that our conclusions are robust, we also used the Void IDentification and Examination toolkit (VIDE) package (Sutter et al. 2015), based on the Zobov void finder (Neyrinck 2008), to determine the voids of the Bolshoi–Planck simulation at $z = 0$.

★ E-mail: tpg2107@columbia.edu (TG); joel@ucsc.edu (JP)

Since the cosmic web has its origin in the gravitational astrophysics of cosmic density and dark energy, a major subject of the many papers investigating the cosmic web has been to identify how the evolution and properties of dark matter haloes are related to their locations within the cosmic web. For example, one area in which there is considerable agreement regards the orientation of the angular momentum of dark matter haloes. In simulations the spin vector of haloes in walls tends to lie in the walls, while the orientation of the spin of haloes in filaments depends on the halo mass and the redshift, with lower mass-halo spins tending to align with the filament while higher mass-halo spins tend to be perpendicular (e.g. Hahn et al. 2007; Aragón-Calvo et al. 2010; Libeskind et al. 2013; Aragon-Calvo & Yang 2014; Wang & Kang 2017). Galaxy observations have not yet yielded a consensus on such spin orientations, although it is true that the spin of the disc of the Milky Way does lie in the Local Wall (e.g. Navarro, Abadi & Steinmetz 2004; Neyrinck 2008; McCall 2014). The edge of this Local Wall is demarcated by a ∼4 Mpc radius ring (filament) of large galaxies that McCall (2014) refers to as the 'Council of Giants', with the Milky Way and Andromeda Galaxy (M31) located near the centre. We show in this paper that the presence of even one such massive halo in such a small wall is rather unusual.

There is strong evidence that properties of dark matter haloes including their masses, and of the galaxies that they host including their masses and luminosities, differ in different web locations (e.g. Eardley et al. 2015, and papers cited there). However, it is important to disentangle the effects of the environmental density and of the web environment. It is the main purpose of this paper to do this for dark matter haloes at $z = 0$, and we will show that the many halo properties that we investigate (except for their orientation) are entirely determined by the environmental density. That is – at least for the definitions of halo properties and environmental density that we adopt – these halo properties at a given environmental density are the same regardless of whether the halo is in a void, wall, or filament. In addition, we do not find evidence at $z = 0$ of special properties of haloes as massive as those of the Milky Way and M31 in walls as small as the Local Wall.

Alonso, Eardley & Peacock (2015) has shown that the mass function of dark matter haloes at the same environmental density is independent of the halo's location in the cosmic T-Web. We study different halo properties in this paper: mass accretion rate, spin parameter, concentration, prolateness, scale factor of last major merger, and scale factor when the halo had half of its $z = 0$ mass. The dependence of these and other halo properties on the halo's environmental density has been discussed in detail in a recent paper Lee et al. (2017a), and this paper extends this analysis to include the location of these haloes within the cosmic web. While the observational results of Tojeiro (2017) show that low-mass haloes in knots are older than haloes of the same mass in voids, appearing to indicate the importance of the cosmic web location, it did not separate location from density, that is, it did not take into account the density effects of Lee et al. (2017a) and Alam et al. (2018). Our results appear to be consistent with observational evidence that properties of nearby galaxies at a given environmental density do not depend on their cosmic web location (Yan, Fan & White 2013; Eardley et al. 2015; Alonso, Hadzhiyska & Strauss 2016; Brouwer et al. 2016). However, there are indications that location in the cosmic web may influence certain properties of galaxies even at the same environmental density, both nearby (e.g. Guo, Tempel & Libeskind 2015) find that SDSS galaxies in filaments have more satellite galaxies than those in other cosmic web locations) and at higher redshifts (e.g. Laigle et al. 2015, 2018).

This paper is organized as follows: In Section 2, we describe the Bolshoi–Planck cosmological simulation and the methods that we use to find and characterize the dark matter haloes, their local densities, and their cosmic web locations. In Section 3, we compare many properties of dark matter haloes in four mass bins as a function of both their environmental density and their locations within the cosmic web, and we find that both the median values and the distributions of these properties are all determined by the environmental density rather than the cosmic web location ; we also study how often haloes as massive as those of the Milky Way and M31 occur in cosmic walls the size of the Local Wall. In Section 4, we summarize and discuss our conclusions. The appendix contains figures that supplement those in the text. Appendix A shows that halo properties are similar in walls of various sizes, at the same cosmic density. Appendix B shows that changing the distance of haloes from the centres of filaments has little effect on the halo properties that we study. Appendix C shows that the distances of haloes in small walls to the centre of their walls have little effect on their angular momentum, which lies in the direction of the planes of their walls.

## 2 METHODS

The following methods were used to study halo properties as a function of density in different web environments in the Bolshoi–Planck simulation with Planck parameters (Klypin et al. 2016; Rodríguez-Puebla et al. 2016): dark matter haloes were found with ROCKSTAR (Behroozi, Weschler & Wu 2013a) and CONSISTENT TREES (Behroozi et al. 2013b); the cosmic dark matter density was Gaussian-smoothed on different length scales (Lee et al. 2017a); and the Bolshoi–Planck simulation was grouped into filaments, walls, and voids using the SpineWeb method (Aragón-Calvo et al. 2010) and the VIDE method (Sutter et al. 2015).

### 2.1 Simulation and halo properties studied

We use the Bolshoi–Planck simulation with $2048^3$ particles in a volume of $(250 \, h^{-1} \, \text{Mpc})^3$ (Klypin et al. 2016; Rodríguez-Puebla et al. 2016). The Bolshoi $N$-body cosmological simulation was made with the Adaptive Refinement Tree code on the Pleiades supercomputer at NASA Ames Research Center. It uses the now-standard $\Lambda$CDM model of the universe and incorporates the results of the Planck Collaboration XXVI (2014) with cosmological parameters: $\Omega_{\Lambda, 0} = 0.693$, $\Omega_{M, 0} = 0.307$, $\Omega_{B, 0} = 0.048$, $h = 0.678$, $n_s = 0.96$, and $\sigma_8 = 0.823$. These cosmological parameters are compatible with the latest Planck results (Planck Collaboration XIII 2016).

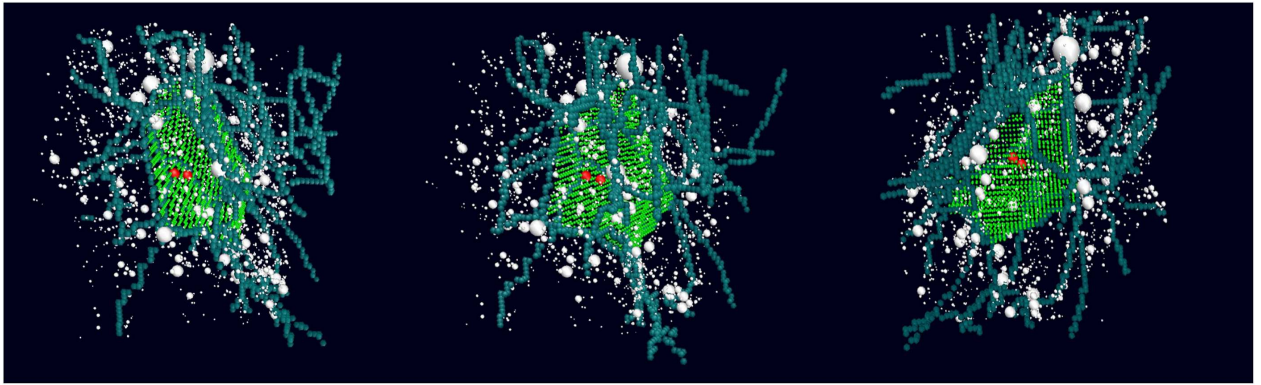

**Figure 1.** A cosmic wall and dark matter haloes visualized in the Bolshoi–Planck simulation. The SpineWeb method (Aragón-Calvo et al. 2010) uses Voronoi tessellation and the discrete watershed transform method to group the cosmic web environment of the $z = 0$ Bolshoi–Planck simulation into cosmic voids, walls, and filaments. Here, we see three views of a wall visualized in green with nearby dark matter haloes shown as white spheres, where the radii of the spheres corresponds to the virial radii of the haloes. The two Milky Way mass haloes near the centre of the wall are shown in red. The nearby cosmic filaments, including those bounding the wall, are shown in teal.

We use the ROCKSTAR (Robust Overdensity Calculation using K-Space Topologically Adaptive Refinement) halo finder (Behroozi et al. 2013a) to identify dark matter haloes in the Bolshoi–Planck simulation. ROCKSTAR is based on the adaptive hierarchical refinement of friends-of-friends groups of particles in six phase-space dimensions plus time. CONSISTENT TREES (Behroozi et al. 2013b) generates merger trees and halo catalogues in a way that ensures consistency of halo mass, position, and velocity across time-steps. This allows it to repair inconsistencies in halo catalogues, and add further information to properties found by ROCKSTAR.[1]

Out of the many halo properties found by ROCKSTAR and CONSISTENT TREES (Rodríguez-Puebla et al. 2016), the main dark matter halo properties that we study are

(i) Specific mass accretion rate (dynamical time averaged) $\dot{M}_{\tau_{dyn}}/M$
(ii) NFW concentration $C_{\rm NFW}$
(iii) Spin parameter $\lambda_{\rm B}$
(iv) Prolateness $P$
(v) Scale factor of the last major merger $a_{\rm LMM}$
(vi) Scale factor when the halo had half of its $z = 0$ mass $a_{M_{1/2}}$

The halo mass accretion rates averaged over a dynamical time are defined as

$$\dot{M}_{\tau_{dyn}} \equiv \left\langle \frac{{\rm d}M_{\rm Vir}}{{\rm d}t} \right\rangle_{\rm dyn} = \frac{M_{\rm Vir}(t) - M_{\rm Vir}(t - t_{\rm dyn})}{t_{\rm dyn}}, \quad (1)$$

where the dynamical time of the halo is $t_{\rm dyn}(z) = [G\Delta_{\rm vir}(z)\rho_{\rm m}(z)]^{-\frac{1}{2}}$, $\rho_{\rm m}(z)$ is the mean matter density at redshift $z$, and $\Delta_{\rm vir}(z)$ is the redshift-dependent virial overdensity (see e.g. Rodríguez-Puebla et al. 2016, fig. 2).

$N$-body simulations have shown that the density profile of most dark matter haloes can be described by the Navarro, Frank & White (Navarro, Frenk & White 1996) profile:

$$\rho_{\rm NFW}(r) = \frac{4\rho_{\rm s}}{(r/R_{\rm s})(1 + r/R_{\rm s})^2}, \quad (2)$$

where the scale radius $R_{\rm s}$ is the radius where the logarithmic slope of the density profile is $-2$. The concentration parameter is defined as the ratio between the virial radius $R_{\rm vir}$ and the scale radius $R_{\rm s}$:

$$C_{\rm NFW} = \frac{R_{\rm vir}}{R_{\rm s}}. \quad (3)$$

Lee et al. (2017b) studied haloes that has suffered significant mass-loss due either to tidal stripping or to relaxation after mergers, and it shows that some such haloes are not well described by equation (2).

The halo spin parameter (Bullock et al. 2001) is defined as

$$\lambda_{\rm B} = \frac{J}{\sqrt{2}M_{\rm vir}V_{\rm vir}R_{\rm vir}}, \quad (4)$$

where $J$ is the total angular momentum of a halo of mass $M_{\rm vir}$, virial velocity $V_{\rm vir}$, and virial radius $R_{\rm vir}$. Lee et al. (2017a) showed that the dependence of $\lambda_{\rm B}$ on density is similar to that of the Peebles spin parameter (Peebles 1969):

$$\lambda_{\rm P} = \frac{J|E|^{1/2}}{GM_{\rm vir}^{5/2}}. \quad (5)$$

The prolateness of the spheroidal dark matter halo (Lee et al. 2017a) is defined as

$$P = 1 - \frac{1}{\sqrt{2}}\left[\left(\frac{b}{a}\right)^2 + \left(\frac{c}{a}\right)^2\right]^{\frac{1}{2}}, \quad (6)$$

where $a \geq b \geq c$ are the lengths of the largest, second largest, and smallest triaxial ellipsoid axes, respectively, determined using the weighted inertia tensor method of Allgood et al. (2006). The prolateness of the simulated haloes ranges from 0 (perfectly spherical) to 1 (maximally elongated, i.e. a needle), with most falling in the range 0.2–0.6 (Lee et al. 2017a).

### 2.2 Density and cosmic web definition

Lee et al. (2017a) implemented a Gaussian smoothing procedure to compute the dark matter density of the full simulation volume smoothed on many different length scales. They convolved the 0.25 $h^{-1}$ Mpc cloud-in-cell (CIC) density cube with a one-dimensional Gaussian kernel applied sequentially along each axis ($x$, $y$, $z$), and smoothed the box on scales of 0.5, 1, 2, 4, 8, and 16 $h^{-1}$ Mpc. Then they added to the information on each halo in the ROCKSTAR halo catalogue the CIC and smoothed density values corresponding to their locations in the simulation volume. The smoothing scale in the figures is chosen to be much larger than the virial radii of the haloes of the corresponding halo mass bins. The smoothing scales in halo mass bins of $\log_{10} M_{\rm vir}/{\rm M}_\odot = 11.20 \pm 0.375$, $11.95 \pm 0.375$, $12.70 \pm 0.375$, and $13.45 \pm 0.375$ were 1, 2, 4, and 8 $h^{-1}$ Mpc, respectively.

---

[1] ROCKSTAR halo catalogues and CONSISTENT TREES merger trees used here are available at http://hipacc.ucsc.edu/Bolshoi/MergerTrees.html, and FOF and BDM halo catalogues are available at https://www.cosmosim.org/cms/simulations/multidark-project/.

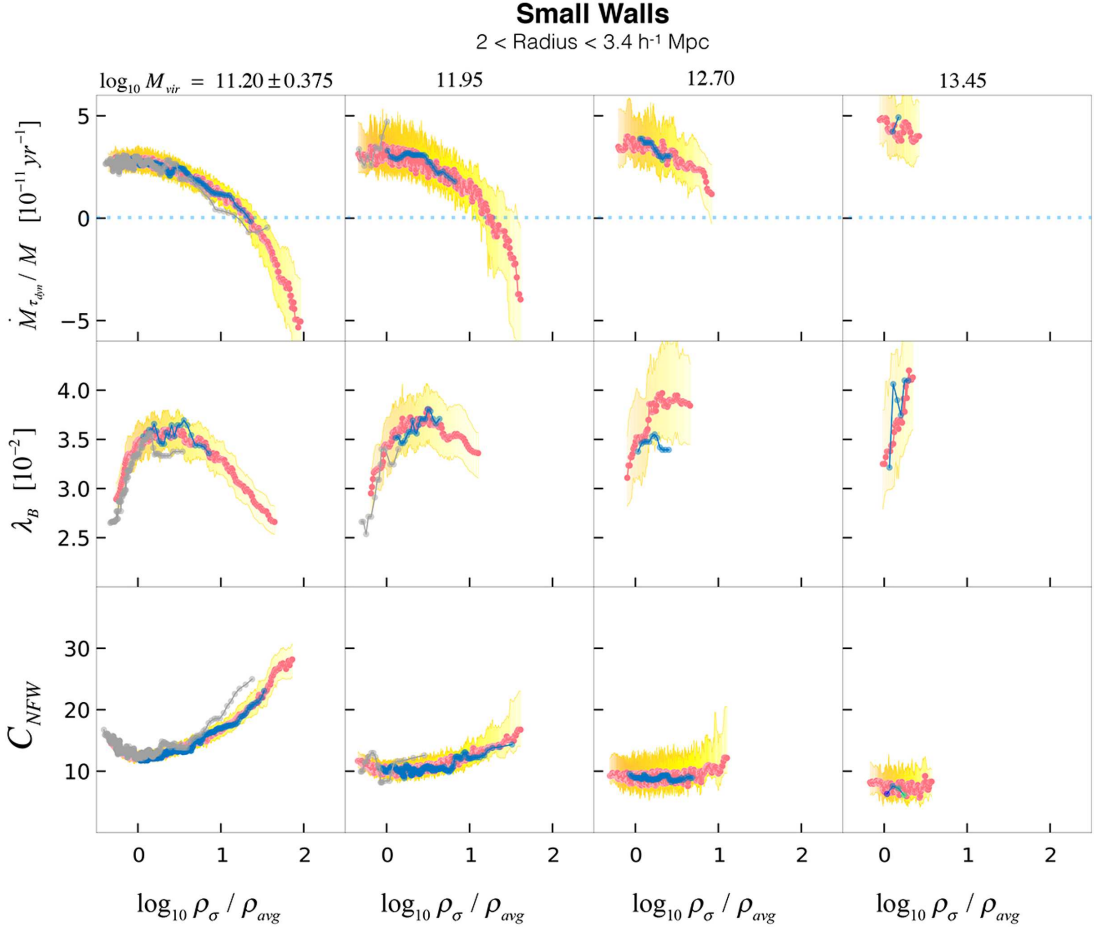

**Figure 2.** Halo properties in all web environments (purple medians and yellow dispersion), small walls (blue medians), and voids (grey medians) as a function of density, where $\rho_\sigma$ is density used on that smoothing scale. *Left to right*: The columns have been split into four mass bins of $\log_{10} M_{\rm vir}/{\rm M}_\odot = 11.20 \pm 0.375$, $11.95 \pm 0.375$, $12.70 \pm 0.375$, and $13.45 \pm 0.375$, with density smoothing scales of $\rho_\sigma$ in these mass bins as 1, 2, 4, and 8 $h^{-1}$ Mpc, respectively. The smoothing scale in each mass bin was chosen so that it is much larger than the haloes in that mass bin, so that the surrounding environment rather than the halo itself mainly determines the density value. *Top to bottom*: The median distribution of specific mass accretion rate, $\lambda_{\rm B}$, and $C_{\rm NFW}$ are plotted versus density in the four mass bins. For these properties, we are following the plots in fig. 5 of Lee et al. (2017a), with the additional step of grouping the haloes into all environments, small walls, and voids via the SpineWeb method of Aragón-Calvo et al. (2010). This allows us to study the median halo properties in the different cosmic web environments. We calculated the median using the moving median method, where we ranked the halo properties according to their density, and calculated the median halo property of a number of haloes as we move towards higher density across each sub-plot. The thick yellow band represents the 5th–95th percentile dispersion of the median of each halo property for all environments. This confidence interval for each halo property is given by $\frac{n}{2} \pm 1.96\sqrt{\frac{n}{4} \mp 1}$, approximated from the binomial distribution, where $n$ is the number of haloes in each bin. Within each sub-plot, we used plot lines and scatter plots to represent the median halo distribution. We note that for all these median plots, the halo properties seem unaffected by the cosmic environment (although there are some tiny fluctuations), and they seem to be controlled instead by the local density. Lastly, we note that for the specific mass accretion rate, in the two lowest mass bins, only haloes in filaments at high densities are losing mass, as indicated by the negative value range of 0 to −5. Haloes at lower densities in filaments, walls, and voids appear to be mainly accreting mass, as indicated by their positive values. The median plots here are for haloes in all environments, small walls, and voids only. In the appendix, Fig. B2 shows the same data as this figure, but with filament radius $D = 0.75h^{-1}$ Mpc instead of $D = 0.25h^{-1}$ Mpc used here.

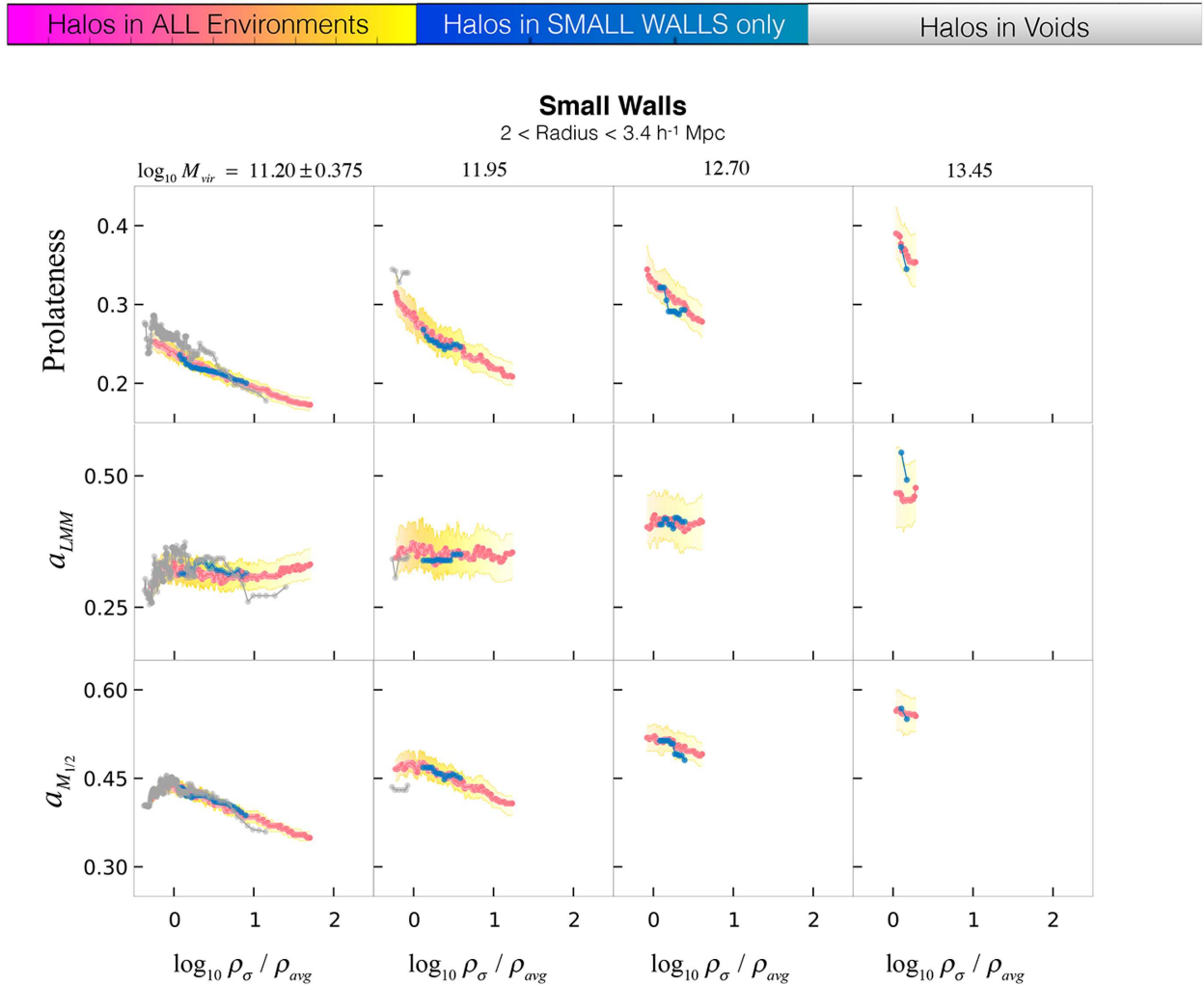

**Figure 3.** Halo properties in all web environments, small walls, and voids as a function of density. *Left to right*: The columns have been split into four mass bins with corresponding smoothing scales, as in Fig. 2. *Top to bottom*: The median distribution of prolateness $P$, scale factor of last major merger $a_{\text{LMM}}$, and the scale factor when the halo had half of its $z = 0$ mass $a_{M_{1/2}}$ are compared with density in different mass bins. For these properties, we are following the plots in figs 9 and 10 of Lee et al. (2017a), with the additional step of dividing the haloes into different cosmic web environments. While these figures in Lee et al. plotted the properties using percentile binning, we instead used the same method as in fig. 5 of Lee et al. in order to see better the different environmental effects on these properties. The median plots here are for haloes in all environments, small walls, and voids only. In the appendix, Figs A1 and A2 show similar plots of median properties in walls of different sizes.

We used two different methods applied to the density field to delineate the cosmic web:

(i) SpineWeb (Aragón-Calvo et al. 2010) for filaments, walls, and small voids (MedianRadius$_{\text{Void}}$ ∼ 4.75 $h^{-1}$ Mpc)

(ii) VIDE (Sutter et al. 2015), which finds larger voids (MedianRadius$_{\text{Void}}$ ∼ 12.5 $h^{-1}$ Mpc),

where MedianRadius$_{\text{Void}}$ is the median radius of all the voids in the simulation via the two different methods.

Aragón-Calvo et al. (2010) implemented SpineWeb using the Watershed Void Finder (Platen, van de Weygaert & Jones 2007) and the topology of the density field to split the cosmic web environment of a simulation into voids, walls, and filaments. SpineWeb is a complete framework for the identification of voids, walls, and filaments by using the watershed transform, in computing the Morse complex, on the cosmic density field. The SpineWeb method invokes the local adjacency properties of the boundaries defined by the watershed segmentation of the field, where the separatrices are classified into walls and filaments. In this procedure, the Delaunay tessellation field estimator (DTFE) method to reconstruct the density field from the spatial particle distribution is applied, where the DTFE procedure produces a self-adaptive volume-filling density field on the basis of the Delaunay tessellation of the point distribution. The identification of these web environments is done on three smoothing scales: 1, 2, and 4 $h^{-1}$ Mpc. For the analysis here, we only use the 2 $h^{-1}$ Mpc smoothing scale, where we used $h = 0.678$, the value used in the Bolshoi–Planck simulation, for all analysis of the simulations.

As McCall (2014) had measured the radius of the Local Wall as ∼ 4 Mpc (2.7 $h^{-1}$ Mpc) with $H_0 = 71.6 \pm 2.9$ km s$^{-1}$ Mpc$^{-1}$, we set a radius range of 2–3.4 $h^{-1}$ Mpc to define walls like our Local Wall as we look for haloes residing in such an environment. We need a smoothing scale that encompasses enough haloes within this radius to do meaningful statistics for small walls like our Local Wall. In addition, we also wanted to do statistics with walls of all sizes, of up to 9.5 $h^{-1}$ Mpc and more, and the smoothing scale would have to encompass enough haloes at these radii as well. As it turns out, the best smoothing scale that gives the widest range is the 2 $h^{-1}$ Mpc smoothing scale. The 4 $h^{-1}$ Mpc smoothing scale yielded too few haloes at the lower wall radius bound below 4 $h^{-1}$ Mpc (most of the haloes have been smoothed out). Similarly, the 1 $h^{-1}$ Mpc smoothing scale yielded too few haloes for walls of radius above 6 $h^{-1}$ Mpc (most of the haloes had been grouped into smaller walls instead of larger walls). With $\sigma_8$ calculated on an 8 $h^{-1}$ Mpc sphere, where the scale is close to linear at this size, a smoothing scale of 4 $h^{-1}$ Mpc would yield not enough rms fluctuations on the number density of galaxies, while that of a 1 $h^{-1}$ Mpc would yield too many fluctuations. Hence, as a compromise, we settled on a 2 $h^{-1}$ Mpc scale to smooth the cosmic web using SpineWeb.

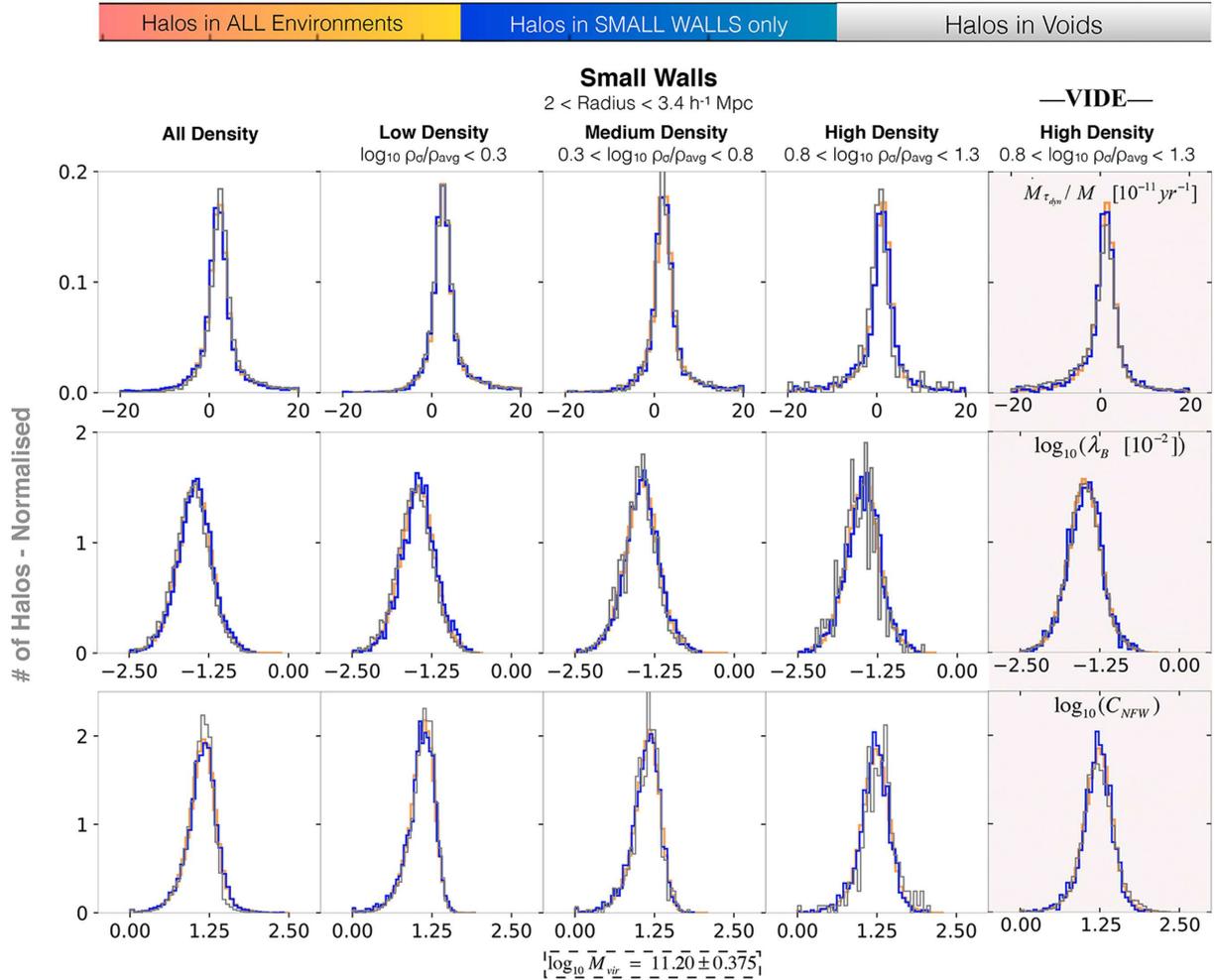

**Figure 4.** Histograms of halo properties in all web environments, small walls, and voids as a function of environmental density, for halo mass $\log_{10} M_{vir}/M_\odot = 11.20 \pm 0.375$ with $\rho_\sigma$ density on the smoothing scale of $1\,h^{-1}$ Mpc. *Left to right*: In addition to plotting all densities, we also split density into three different ranges to explore the effect of low, medium, and high densities on the halo properties, and we plot at the right the histograms for large densities in large voids found using VIDE. *Top to bottom*: Specific mass accretion rate $\dot{M}/M$, $\lambda_B$, and $C_{NFW}$ are compared with density in small walls (radius 2–3.4 $h^{-1}$ Mpc) (in blue), voids of all sizes (grey), and environments of all sizes (orange). $\lambda_B$ and $C_{NFW}$ follow a lognormal distribution, while the accretion rate follows a Gaussian distribution as the variables that make up its properties are random. The curve-fitting is detailed in Fig. 6 of this paper. To create the bins, we cut off the tail-ends of each histogram at the ranges shown above, as there are not enough haloes in the tails for meaningful statistics. Then, we split the distribution into 40 bins per property. We note that for the accretion rate, haloes in all environments, small walls, and voids are losing mass (negative values) as well as gaining mass (positive values) with no distinction between environments.

Using SpineWeb, Fig. 1 shows visualizations of a cosmic wall like our Local Wall containing two Milky Way mass haloes (i.e. with mass $\sim 10^{12}$ M$_\odot$), viewed from three different directions. Here, the white spheres are dark matter haloes and the two red spheres represent the Milky Way mass haloes in this wall. The green dots show the wall, and the teal spheres represent the nearby filaments, including the filaments bounding the wall.

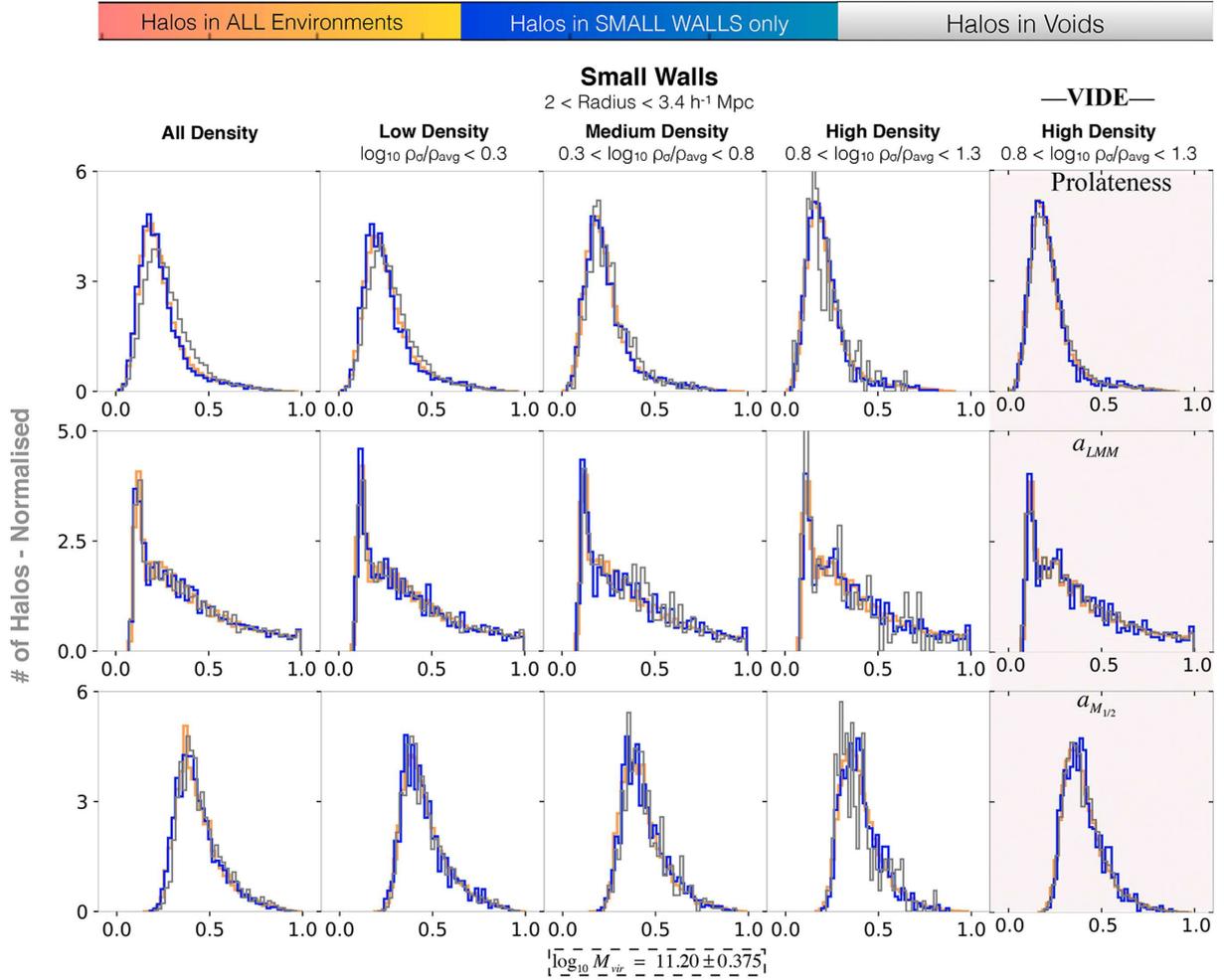

**Figure 5.** Histograms of halo properties in all web environments, small walls, and voids as a function of density smoothed on a scale of $1\,h^{-1}$ Mpc, for halo mass $\log_{10} M_{\rm vir}/{\rm M}_\odot = 11.20 \pm 0.375$. *Left to right*: As in Fig. 4. *Top to bottom*: The full distribution of prolateness $P$, scale factor of last major merger $a_{\rm LMM}$, and scale factor when the halo had half of its $z=0$ mass $a_{M_{1/2}}$ are compared with density in small walls (radius 2–3.4 $h^{-1}$ Mpc) (in blue), voids of all sizes (grey), and environments of all sizes (orange). For the scale factor of the last major merger, there is no lognormal distribution, as the histogram indicates the time when the merger occurs. There is a drop-off after 0.0 as every halo would have been formed by mergers in the early universe. However, as CONSISTENT TREES trace the dark matter particle into the past, the tracing on the merger trees becomes unreliable for $z \gtrsim 10$ or when the haloes have fewer than 50 dark matter particles. We note that while there are some tiny deviations in the median plots of Figs 2 and 3, the full distribution here reveals that the tiny fluctuations are negligible, and are not a result of the cosmic web environment.

The SpineWeb filaments and walls have the thickness of single voxels, cubes of side $0.25\,h^{-1}$ Mpc, in the Bolshoi–Planck simulation. We introduce a parameter $D$, the distance to filaments. We define this as a radius around a filament, forming a cylinder within which the haloes are defined to be within a filament. We use $D = 0.25\,h^{-1}$ Mpc for this paper. (We also tried $D = 0.75\,h^{-1}$ Mpc and we found that the results presented below are very similar. The similarity between using $D = 0.25\,h^{-1}$ Mpc and $D = 0.75\,h^{-1}$ Mpc is shown in Figs B1 and B2.) We similarly assign haloes to walls that are within a distance $0.75\,h^{-1}$ Mpc, and not assigned to filaments.

To find and characterize the properties of larger voids in the Bolshoi–Planck simulation, we used the VIDE (Sutter et al. 2015) method, which similarly calculated a Voronoi tessellation for estimating the density field, and performed a watershed transform to construct voids.

### 2.3 Analysis

Lee et al. (2017a) determined how halo properties including the specific mass accretion rate, $\lambda_{\rm B}$, and $C_{\rm NFW}$ depend on density for haloes in all cosmic web locations. We are extending that work by grouping the haloes into different web environments of filaments, walls, and voids, and studying the effects of density on halo properties in those web environments. We did not look separately for halo properties in filaments as 62 per cent of haloes are already within a cylindrical radius of $D = 0.25\,h^{-1}$ Mpc of the filaments, which

**Curve-fitting to specific accretion rate for all densities**

| Accretion Rate [$10^{-11}$ yr$^{-1}$] | Median | Mean | Standard Deviation | Model |
|---|---|---|---|---|
| Small Walls | 1.542 | 2.024 | 2.392 | Gaussian |
| Medium Walls | 2.485 | 2.297 | 2.349 | Gaussian |
| Large Walls | 2.524 | 2.241 | 2.368 | Gaussian |
| X-Large Walls | 2.508 | 2.168 | 2.338 | Gaussian |
| All | 2.500 | 2.093 | 2.400 | Gaussian |
| Voids | 2.497 | 2.412 | 2.322 | Gaussian |

**Curve-fitting to $\log_{10} \lambda_B$ for all densities**

| $\log_{10} \lambda_B$ [$10^{-2}$] | Median | Mean | Standard Deviation | Model |
|---|---|---|---|---|
| Small Walls | -1.428 | -1.461 | 0.264 | Lognormal |
| Medium Walls | -1.511 | -1.463 | 0.262 | Lognormal |
| Large Walls | -1.481 | -1.456 | 0.263 | Lognormal |
| X-Large Walls | -1.470 | -1.461 | 0.263 | Lognormal |
| All | -1.457 | -1.469 | 0.264 | Lognormal |
| Voids | -1.460 | -1.506 | 0.268 | Lognormal |

**Curve-fitting to $\log_{10} C_{\rm NFW}$ for all densities**

| $\log_{10} C_{\rm NFW}$ | Median | Mean | Standard Deviation | Model |
|---|---|---|---|---|
| Small Walls | 1.181 | 1.155 | 0.210 | Lognormal |
| Medium Walls | 1.083 | 1.120 | 0.203 | Lognormal |
| Large Walls | 1.146 | 1.129 | 0.203 | Lognormal |
| X-Large Walls | 1.168 | 1.137 | 0.205 | Lognormal |
| All | 1.154 | 1.147 | 0.206 | Lognormal |
| Voids | 1.146 | 1.142 | 0.181 | Lognormal |

**Figure 6.** Distributions of specific halo mass accretion rate $\dot{M}/M$, spin parameter $\lambda_B$, and NFW concentration $C_{\rm NFW}$ of haloes within $< 0.75 \, h^{-1}$ Mpc away from walls and $> 0.25 \, h^{-1}$ Mpc away from filaments. Corresponding to Fig. 4, we fitted a Gaussian distribution to the specific accretion rate, and lognormal distributions to $\lambda_B$ and $C_{\rm NFW}$. As a result of the similar values between the fitting parameters in different cosmic web environments (voids, walls, and all web environments), we can conclude that the entire distribution of properties in different web environments is similar, and that the cosmic web environment does not appear to affect the distribution of halo properties. We note that the median mass accretion rate for the small walls is lower than for other size walls; we explore this in Fig. B1.

is to say that the haloes in all environment is dominated by those in filaments. Doing a separate analysis for haloes in filaments would not change the conclusion that we will show later in this paper. Instead, we analysed the haloes separated into these three cosmic environments: all web environments, walls, and voids.

Note that we used the exact same haloes as Lee et al. (2017a), where high-mass haloes were randomly removed in each mass bin in order to remove dependence of density on mass and get a flat mass distribution in each mass bin. This was done so that the halo properties would be dependent on density alone. Our results below show the effects of web location and density on haloes of the same mass.

## 3 RESULTS

We split the presentation of the results into two subsections:

(i) **General results**: We look at the halo properties of dark matter haloes across all filaments, walls, and voids. The results here can

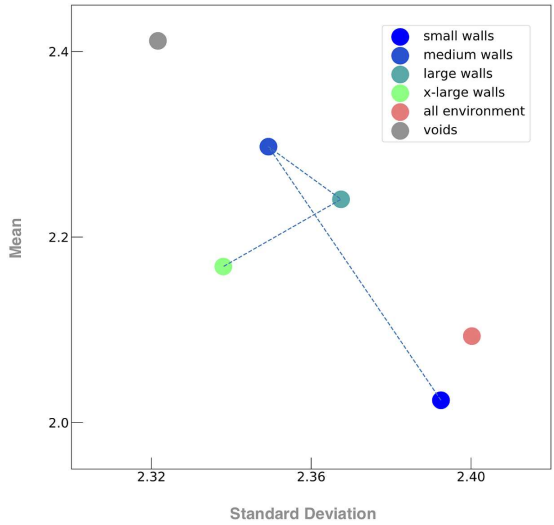

**Figure 7.** Standard deviation and mean of the Gaussian distribution of the specific mass accretion rate $\dot{M}/M$ as a function of environment. The dotted line refers to the distribution under gradual increase in size of the walls. The random walk would seem to indicate that the sizes of walls has no effect on the distribution of the halo property and the close values of the parameters would indicate that different web environments have little effect on the distribution as well.

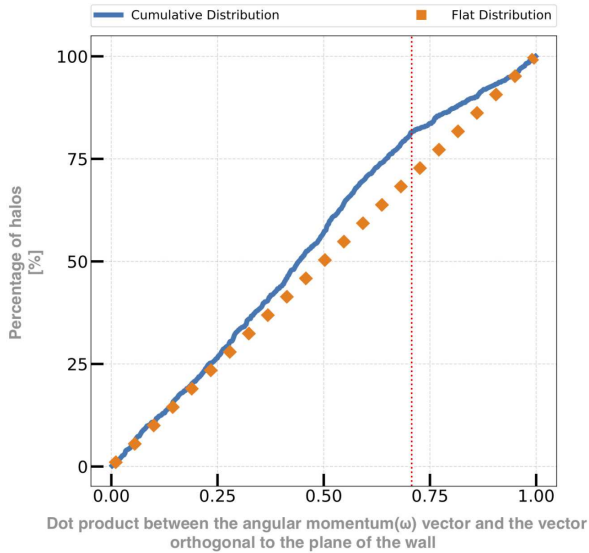

**Figure 8.** The cumulative distribution of the angle between the angular momentum vector ($\omega$) of wall haloes and a vector orthogonal to the plane of the wall, shown in blue. We additionally show a flat distribution for comparison in orange. The dot product of a $\omega$ vector in the plane of the wall with a vector orthogonal to the plane will be 0. We see that most of the haloes make a large angle/small dot product with their $\omega$ vector and the vector orthogonal to the plane of the wall, as evidenced by the dot product at $45°$ $\sim 0.707$ (the vertical red line), showing that about 80 per cent of the haloes have angular momentum vectors within $45°$ of the plane of the wall. To compute the mean angle that all the haloes make with their $\omega$ vector and the plane of the wall, we first computed the mean dot product for all 1099 dark matter haloes with their respective perpendicular vector to wall. This mean dot product is 0.45, which indicates a mean angle of $30°$ that the $\omega$ vector makes with the plane of the wall. Our results are consistent with those of Aragón-Calvo et al. (2010).

be generalized across all distinct haloes and are not confined to just those of our own Local Wall.

(ii) **Local Wall results**: We then go on to look in more details at walls, in particular walls like our Local Wall, as we are particularly interested to see if haloes in walls like our own Local Wall have properties different from other web environments.

### 3.1 General results

For the general results, we will be looking at the following:

(i) The plots of median distributions of halo properties in different cosmic web environments by the SpineWeb method.
(ii) The histograms of the full distribution of these halo properties in different cosmic web environments by the SpineWeb method.
(iii) The halo properties in larger walls found using the VIDE code.

*3.1.1 Plots of SpineWeb halo properties*

In Figs 2 and 3, we present the plots of halo properties (specific mass accretion rate, spin parameter, NFW concentration, prolateness, scale factor at last major merger, and the half mass scale factor) against density for haloes in all types of environment, in small (2–3.4 $h^{-1}$ Mpc) walls, and in voids of all sizes, across various dark matter halo mass bins. We define 'small' walls as having size 2–3.4 $h^{-1}$ Mpc in order to determine properties of haloes of walls like our Local Wall, as McCall (2014) has found the edge of our Local Wall to be at an $\sim 2.7\,h^{-1}$ Mpc radius ring (filament) of large galaxies. The effects of larger walls on halo properties are found in the appendix. The density in all cosmic environments ranges from $\log_{10} \rho_\sigma/\rho_{\text{ave}} = -0.5$ to 2.

It should be noted that only the first two smaller mass bins yielded enough haloes to allow meaningful statistical interpretation, particularly for the case of voids where the two higher mass bins yielded few haloes, with the highest mass bin having almost no haloes in voids.

To balance each mass bin to have a flat mass–density relation, Lee et al. (2017a) did a two-dimensional sub-binning by halo mass and a given local density parameter for each given mass bin, then randomly eliminated haloes from appropriate sub-bins to force approximately equivalent mass distributions for each density sub-bin. We used the haloes in these mass bins, and then plotted the median of each halo property as a function of density for each mass bin.

In addition, we show the lower and upper bound on the 95 per cent confidence interval of the median, which can be seen in the plots as the thick yellow band. Moreover, we performed a smoothing for the plots using the moving median, where we took the median value of the halo properties at every few points as we move from left to right across the density axis, in order to remove noise.

We see some tiny deviations of halo properties in the different web location of all environments, walls, and voids, particularly as the log-density increases from 0.5 onwards. For example, in Fig. 2, we see that the accretion rate in voids appears to fall below that in all environments around log-density = 0.8. Conversely, we see that for NFW concentration, the haloes in voids appear to have a larger concentration than those in all environments. However, these deviations could be misleading, particularly because there are fewer haloes towards the high-density end of each plot, especially for voids. As a result, the deviations could have just arisen due to insufficient data. In order to see if there is any effect that the cosmic web environment has on these properties, we need to look at the full distributions, instead of the medians, where some of the information might have been smoothed away. For these full histograms, see Figs 4 and 5. It should be stressed that the wall results in Figs 2–5 are for the geometric environments of small walls (2–3.4 $h^{-1}$ Mpc), filaments, and voids, as we are interested in knowing whether our own Local Wall has any peculiarity affecting these halo properties. These figures are discussed further in Section 3.2.2.

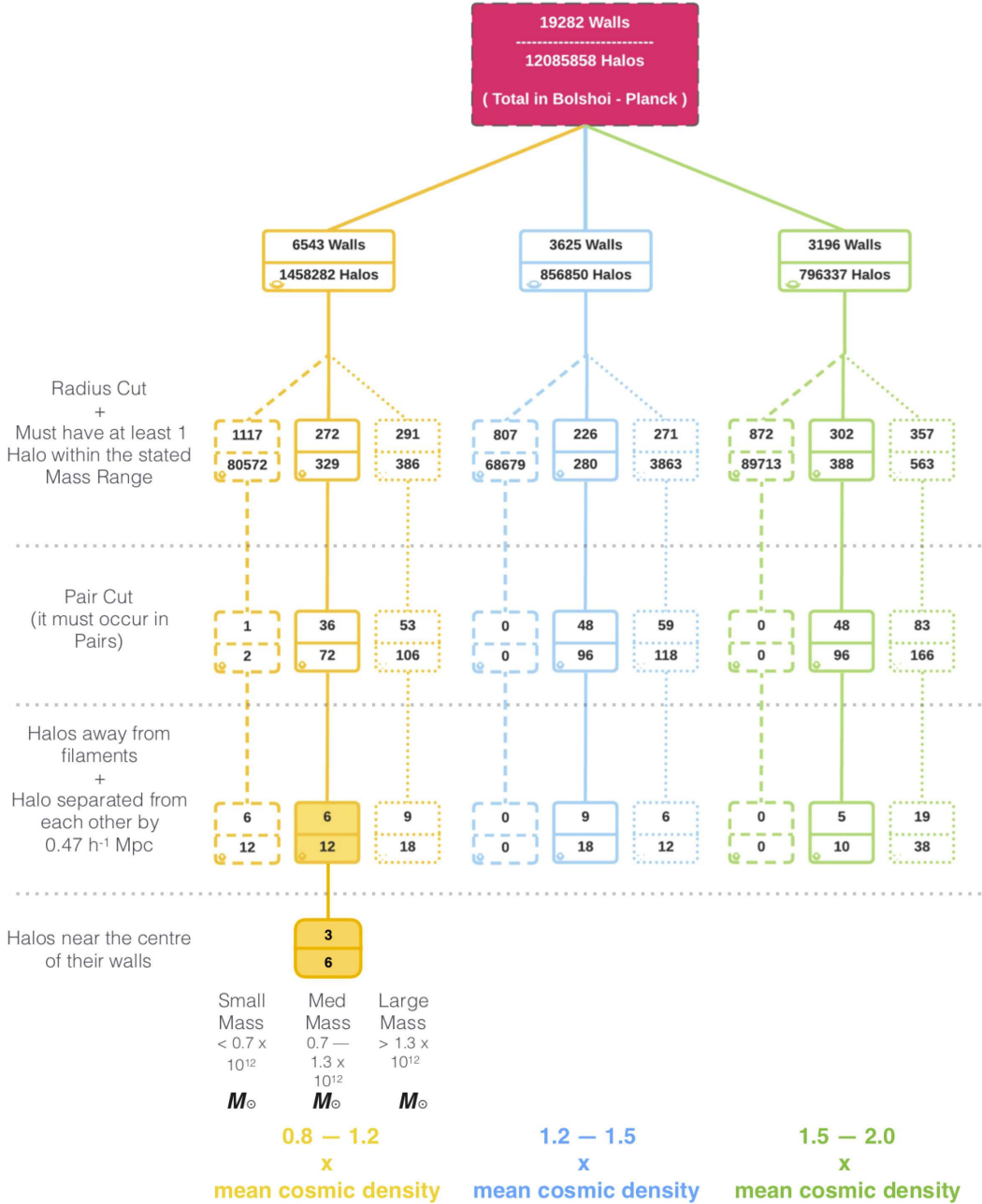

**Figure 9.** Flowchart of the cuts made to find halo pairs like the Milky Way Galaxy and Andromeda. Our Local Wall is found to have a radius of 2.7 $h^{-1}$ Mpc, while the masses of our Milky Way Galaxy and Andromeda are thought to be about 0.8–1.2 × $10^{12}$ M$_\odot$. The cosmic density of the local (∼10 Mpc radius) environment is approximately equal to the average density of the universe (Klypin et al. 2015). By using the haloes of Lee et al. (2017a), we found the density of the walls to be 1.02, 1.35, and 1.75, respectively, for small haloes, medium haloes, and large haloes, all calculated with the $\rho_\sigma$ density on the smoothing scale of 8 $h^{-1}$ Mpc. As SpineWeb assigns walls that are only a voxel thick (corresponding to only 0.25 $h^{-1}$ Mpc), we set a structural reassignment criterion of assigning haloes to walls if they are >0.25 $h^{-1}$ Mpc from filaments to more accurately reflect criteria found in surveys. Second to last, we wanted the wall to contain a pair of dark matter haloes (Milky Way + Andromeda), whose distance is less than about 0.47 $h^{-1}$ Mpc apart. We are left with 6 walls out of an initial 19 282 Walls, which give us 0.06 per cent of all walls in the Bolshoi–Planck, which are similar to our Local Wall. This makes configurations like our Local Wall very rare. In addition, when we looked closely at these 6 walls/12 haloes, we found that 2 of the haloes were actually sub-haloes. There are thus only 5 walls/10 haloes left in the Bolshoi–Planck simulation whose properties are like those of the Local Wall. Lastly, if we further used the map of McCall (2014) to restrict haloes to the centre of their walls like our Local Wall, we end up with just 3 walls/6 haloes left, which gives us 0.03 per cent of all walls in the Bolshoi–Planck simulation.

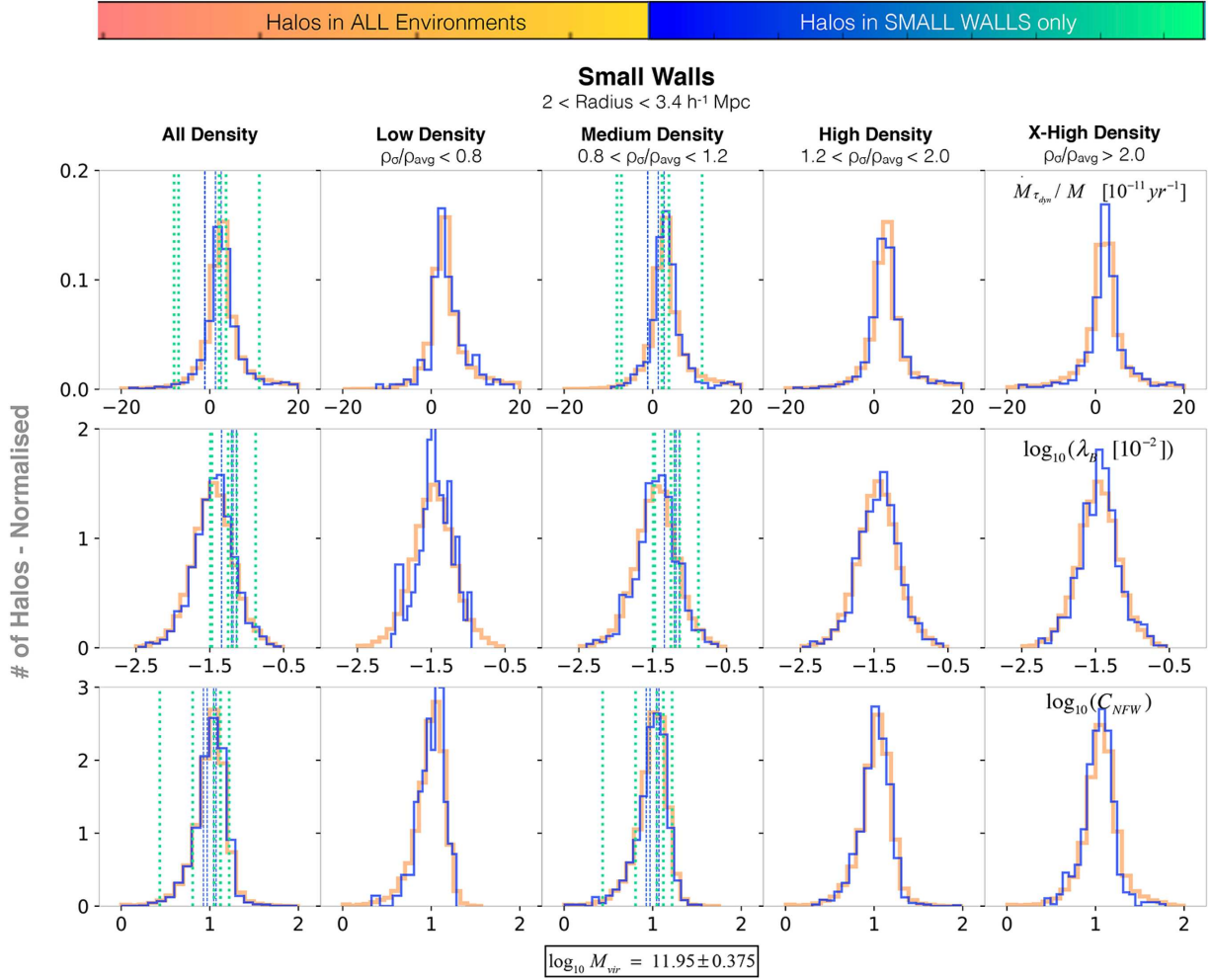

**Figure 10.** Histograms of halo properties in all web environments and small walls as a function of environmental density, for halo mass $\log_{10} M_{\rm vir}/M_\odot = 11.95 \pm 0.375$ with $\rho_\sigma$ density on the smoothing scale of $8\,h^{-1}$ Mpc. *Left to right*: In addition to plotting all densities, we also split density into four different ranges to explore the effect of low, medium, high, and extra-high densities on the halo properties. The six vertical green dotted lines ($<1.75\,h^{-1}$ Mpc from the centre of mass of the walls) along with four vertical blue dashed lines (otherwise) make up the 10 haloes most like our own Milky Way Galaxy and Andromeda. *Top to bottom*: Specific accretion rate $\dot{M}/M$, $\lambda_B$, and $C_{\rm NFW}$ are compared with density in small walls (radius 2–3.4 $h^{-1}$ Mpc) (in blue) and environments of all sizes (orange). Unlike the histograms of Figs 4 and 5, we did not plot the halo properties in voids here as there are too few haloes in voids at the stated mass range calculated at the stated smoothing scale to do meaningful statistics: the number of haloes in voids in all density ranges here is just 59, while the number of haloes in all environment in the same density range is 74 121, while that of walls is 2401. It should be noted these numbers in the all density range refer to haloes in their respective environments after cuts made to their masses ($\log_{10} M_{\rm vir}/M_\odot = 11.95 \pm 0.375$), after cuts made to the radius of the wall (2–3.4 $h^{-1}$ Mpc), after cuts made to being away from filament ($>0.25\,h^{-1}$), and after cuts of reassigning haloes further away in voids to walls instead ($>0.75\,h^{-1}$ from walls). The first 3 cuts correspond to the first 3 criteria found in Section 3.2.2, and while we used $<0.75\,h^{-1}$ Mpc for haloes distance to walls to split the environment in the histograms, we did not use this criterion for haloes in our Local Wall (the vertical lines), as we wanted to look for haloes exactly in the plane of the wall there (McCall 2014). Also unlike the histograms of Figs 4 and 5, the histograms here are not logged, due to the low overall density of haloes measured at this smoothing scale. The histograms here corresponds to the mass range in the second column of Fig. 2, although it is calculated at the smoothing scale of 8 $h^{-1}$ Mpc, and not 2 $h^{-1}$ Mpc of Fig. 2. After additional cuts made to the number of haloes listed as items v–vii in Section 3.2.2, we found 10 haloes that are like those of the Milky Way Galaxy and Andromeda (the methods that we used to select them are found in full in 3.2.2, and illustrated with a flowchart in Fig. 9), which we have illustrated above as blue dotted lines. We note here that these 10 haloes fall roughly in the median of these histograms with known Gaussian ($\dot{M}/M$) and lognormal distributions ($\lambda_B$ and $C_{\rm NFW}$), indicating that these halo properties of the Milky Way galaxy and Andromeda in our Local Wall are not peculiar.

In order to account for halo properties in different-sized walls, we made similar plots for walls of different sizes in Figs A1 and A2 in the appendix. For the different-sized walls, we see that

(i) small walls (2–3.4 $h^{-1}$ Mpc),
(ii) medium walls (3.4–6.8 $h^{-1}$ Mpc),
(iii) large walls (6.8–9.5 $h^{-1}$ Mpc),
(iv) extra-large walls (> 9.5 $h^{-1}$ Mpc),

where the number of haloes per wall that we ended up analysing is 27 405, 8185, 1910, and 324, respectively, the halo properties mainly fall within the lower and upper bound on the 95 per cent confidence interval of the median. For the full distribution of halo properties, we refer to the histograms of all walls in Figs A3 and A4 in the appendix. We note that there do not appear to be any real deviations in halo properties between the different-sized walls when compared in the same mass bin. Thus, we find that halo properties in all web environments, small walls, and voids are essentially the same for walls of all sizes at the same environmental density.

### 3.1.2 Histograms of SpineWeb halo properties

In addition to the median distribution plotted in Figs 2 and 3, we also studied the full distribution of the halo properties. In Figs 4 and 5, we present the histograms of these full distributions of the halo properties in small walls, voids, and all environments. We split the histograms up into regions of low to high density in order to study the overall effect of densities on the distribution. As there are not enough haloes in the tails of the histogram for meaningful statistics, we concentrate on the peak of the full distribution by cutting cut off the tails of each histogram in the range of the *x*-axis shown in the diagrams. We then split each range into 40 bins. The density ranges from $\log_{10}\rho_\sigma/\rho_{ave} = -0.5$ to 2.

To see the real limitation of the median plots of Figs 2 and 3, we shall point out that the entire *y*-axis range of $-5$ to $5 \times 10^{-11}$ yr$^{-1}$ for the specific mass accretion rate of Fig. 2 fits into just the median 10 bins out of 40 bins of the full distribution of the accretion rate of Fig. 4. This full distribution of the specific mass accretion rate, as well as for the other halo properties of $\lambda_B$ and $C_{NFW}$, shows us that any tiny deviations that arise out of the differences in the distribution of halo properties in filaments, small walls, and voids are almost entirely negligible.

To quantify this, we made curve-fitting plots for the specific mass accretion rate $\dot{M}/M$, $\lambda_B$, and $C_{NFW}$. In Fig. 6, we fitted the accretion rate to a Gaussian distribution, and $\lambda_B$ and $C_{NFW}$ to lognormal distributions (as in Rodríguez-Puebla et al. 2016).

In order to create a curve through our data points, we took the *y*-axis quantity of each halo property bin and treated that as a scattered point through which we drew a data curve. We then obtained the modelled curve via NUMPY CurveFit, where the best-fitting parameters (mean and standard deviation) were acquired by getting the best-fitting curve (the model) given the histogram (our data).

In Fig. 7, we plotted the standard deviation and mean of the specific mass accretion rate $\dot{M}/M$ in various cosmic environments. The dotted line in the plot shows the distribution under gradual increase in size of the walls, and while these parameters for walls appear to be sandwiched between those for voids and all cosmic web environments, the effect on the general Gaussian distribution of the histogram of $\dot{M}/M$ is tiny, showing that the cosmic web environment has little effect on this halo property.

As a result of the similar values between the fitting parameters in different cosmic web environments (voids, walls, and all web environments), we can conclude that the entire distribution of properties in different web environments is similar, and that the cosmic environment does not appear to affect the distribution of halo properties that we studied.

### 3.1.3 Void halo properties using VIDE

The SpineWeb method splits the simulation into voids that are mostly rather small, with an average void radius of ∼5 Mpc. For large voids, we used the VIDE method (Sutter et al. 2015), where the voids are similar in size to those found by VIDE on SDSS Release 9. With this VIDE method, the average void radius is ∼13 Mpc. In the last column of Figs 4 and 5, we find that even with these larger voids, the histograms with various sized walls remain similar to those with smaller voids using the SpineWeb method. We note that the larger voids of VIDE can host more haloes, and as a result their full distribution void histograms are less fragmented than those using SpineWeb on smaller sized walls. We find that using the VIDE method produces histograms of properties that are a closer fit with one another across different cosmic web environments than using the SpineWeb method. For example, in Fig. 3, we note that the haloes in voids appear to produce a higher prolateness in the first and second mass bin. We note that for the full distribution of the histogram of prolateness in Fig. 5, the prolateness of haloes in voids is less than the haloes in all environment and walls. We also note that with our $D = 0.25 h^{-1}$ Mpc filament radius parameter, haloes in voids only account for 2 per cent of all haloes. However, using the VIDE method we obtain larger voids and 30 per cent of haloes reside in them. We see the result of this higher percentage in the last column of Fig. 5, where there are no differences between the prolateness of haloes in all environments, walls, and voids. Hence, we conclude that for full distribution using the VIDE method, the cosmic web environment does not significantly affect the halo properties.

## 3.2 Local Wall

### 3.2.1 Angular momentum orientation in local walls

Aragon-Calvo & Szalay (2013) have found that the SpineWeb method produced dark matter haloes in walls whose angular momentum axes lie close to the plane of the wall. We wanted to check that this property still holds true when the SpineWeb is applied to the Bolshoi–Planck simulation. To verify this, we first look for walls like our own Local Wall by applying the following cuts to the initial 19 281 walls of the Bolshoi–Planck simulation:

(i) Radius (wall): 2–3.4 $h^{-1}$ Mpc
(ii) Mass (haloes): 0.7–1.3 $\times 10^{12}$ M$_\odot$
(iii) Distance of halo to filaments: >0.25 $h^{-1}$ Mpc

It should be noted that these three cuts do not include a requirement that wall haloes be <0.75 $h^{-1}$ Mpc away from walls, which is a criterion we used elsewhere in the paper to group haloes to different environment. We did not use this criteria in the analysis of the Local Wall as we wanted to look for haloes exactly in the plane of the wall here (McCall 2014).

These three criteria above bring the number of walls down to 594, and the number of dark matter haloes in these walls down to 702. The first two cuts give us the haloes the size of the Milky Way in a wall the size of our Local Wall. The remaining cut (iii) is used throughout this paper as part of the criteria to assign haloes to their respective walls, as SpineWeb only assigns haloes to filaments and walls that are $0.25\,h^{-1}$ Mpc thick around the haloes (as the size of a voxel by SpineWeb is only $0.25\,h^{-1}$ Mpc on each side), although the length and planar dimensions are much longer and larger. By setting the distance of halo to filaments to be $>0.25\,h^{-1}$ Mpc, we are using assignment criteria that more accurately reflects the cosmic environment found in surveys in terms of locating wall haloes away from filaments, while keeping those haloes located exactly in the voxels of their wall. Using the distance of halo to filaments to be $>0.25\,h^{-1}$ Mpc has the additional effect of pre-selecting haloes to lie closer to the centre of their walls for small walls, as $>0.25\,h^{-1}$ Mpc is large compared with the radii 2–3.4 $h^{-1}$ Mpc of small walls. However, these haloes in the wall might still be too close to filaments if we just set the distance of halo to filaments to be just $>0.25\,h^{-1}$ Mpc, so we explored limiting the calculation of angular momentum to just the centre of these walls in Appendix C, and the results we found there are qualitatively similar to the results we will show later in this section.

These three cuts in this section give us enough walls (594) to still do meaningful statistics. For the 594 walls that are similar enough to our Local Wall, to determine whether the angular momenta $\omega$ of these dark matter haloes in the walls lie within the plane of those wall, we applied the following procedure:

(i) We found a pair of nearby vectors that lie in a locally flat region around the dark matter halo, by choosing the few pixels closest to the halo that are in the plane of the wall and drawing a line between those pixels to form a vector.

(ii) The cross product of these vectors is orthogonal to the plane of the wall.

(iii) Finally, we calculated the dot product between this orthogonal vector and the $\omega$ vector.

If $\omega$ vector were lying exactly in the plane of the wall, then its dot product with the orthogonal vector should be 0. We expect the angular momentum to lie close to the plane of the wall, so the angle that the $\omega$ vector makes with the plane of the wall should be less than 45°, and the angle that it makes with the orthogonal vector should be more than 45° (their dot product tending towards 0 in Fig. 8). We applied the above dot product procedure to all the dark matter haloes that are near the centres of those 594 walls. Fig. 8 shows the cumulative distribution of these angles. Then we took the mean of those dot products, and obtained the following results:

(i) Of all dark matter haloes near the centres of 594 walls, the mean of the dot products between their $\omega$ vector and the orthogonal vector = 0.452. [When calculating these orthogonal vectors to the wall, we calculated using 160, 80, and then 20 voxels around the dark matter haloes. Correspondingly, for these numbers of voxels around the haloes, we found the dot product (via the method stated above) to be 0.4544, 0.4524, and 0.4523, which is to say that there is no significant changes when using different sets of neighbouring voxels.]

(ii) As a result, the mean angle ($\omega$ vector) = $\sim$60° to the vector that is orthogonal to the wall.

(iii) The mean angle of $\omega$ vector with respect to the plane wall is $\sim$30°.

We can see these results visualized in the cumulative distribution of dot products (between the $\omega$ vector and the orthogonal vector) in Fig. 8. The orange dots are an idealized flat distribution of dot products, with the true cumulative distribution represented in blue. We can see that 50 per cent of the haloes corresponds to the mean dot product of $\sim$0.45, which is $\sim$30° to the plane of the wall. Moreover, if the $\omega$ vector were lying at $\sim$45° to plane of the wall, then its dot product with the orthogonal vector will be $\frac{1}{\sqrt{2}}$, or 0.707, as indicated in the figure by the vertical red dotted line. This corresponds to nearly 80 per cent of all haloes making an angle of less than $\sim$45° with the plane of the wall. It should be noted, however, that for a perfectly flat, isotropic distribution, 71 per cent of all haloes would have angles smaller than 45°, and the mean angle with respect to the wall would be 33°.

Hence, our results here confirm that for walls similar to our own Local Wall, using the SpineWeb method, the mean angle between the halo angular momentum and the plane of the wall is about 30 deg, i.e. haloes have a slight tendency to align in the direction of their angular momentum with the plane of the wall, although the dispersion is high.

### 3.2.2 Halo properties in the local walls

McCall (2014) found that our Milky Way Galaxy and the Andromeda Galaxy lie near the centre of a wall of $\sim$4 Mpc in radius. By grouping haloes in the $z = 0$ Bolshoi–Planck simulation into filaments, walls, and voids, we can look for walls that most closely resemble our own Local Wall and study the properties of the haloes residing in such walls. If our own Local Wall has special halo properties, that is, if the halo properties in such Local Wall do not fall generally within the median of their known distribution, then that would suggest that the near-field cosmology of our own Local Group is peculiar. The properties that we can find observationally of the Milky Way and its dark matter halo could not then be generalized for all galaxies and their halo companions.

To find walls like our own with the Milky Way and Andromeda galaxies in it, we performed the following cuts (with $h = 0.7$):

(i) Radius (wall): 2–3.4 $h^{-1}$ Mpc
(ii) Mass (haloes): $0.7$–$1.3 \times 10^{12}$ M$_\odot$
(iii) Distance (halo to filament): $> 0.25\,h^{-1}$ Mpc
(iv) Density (wall): $0.8-1.2 \times$ average cosmic density
(v) Must occur in pairs of distinct haloes
(vi) Distance (between pairs): $<0.47\,h^{-1}$ Mpc
(vii) Distance from centre of wall: $<1.75\,h^{-1}$ Mpc

The first three items were used in Section 3.2.1 when we were determining the angular momentum. For this section, we extended these three cuts to include cosmic density (based on a survey of galaxies in the local region within about 10 Mpc (Klypin et al. 2015), and that the pair of haloes in the wall must be within 0.7 Mpc of each other, just like in our Local Wall. We present a flowchart of the different cuts we made to the simulation in Fig. 9. From an initial 19 281 walls, we are left with just 6 walls/12 haloes that most resemble our Local Wall. On closer inspection, 2 of the haloes were really just sub-haloes of a group, unlike the Milky Way and Andromeda galaxies which are in separate distinct haloes (being distinct means that though they are gravitationally bound together, their virial radii do not overlap). So we end up with just 5 walls/10 haloes. We took these remaining 10 haloes that are most like our own Milky Way and Andromeda, and then further limit them to the centre of their walls, as mapped out by McCall (2014). This gives us just 3 walls/6 haloes that are like ours in the Local Wall. Hence, we conclude that our Local Wall accounts for only 0.03 per cent of all walls in the Bolshoi–Planck simulation.

We note here that although we had put in several constraints to find walls like our own with the Milky Way Galaxy and Andromeda in it, our Local Wall does not need so many restrictions for it to be considered rare, as most of the ∼12 million haloes do not reside in walls; indeed, only ∼3 million are in walls as haloes are mainly in filaments. Moreover, most haloes are not found in the density range of 0.8–1.2 times the volume-averaged cosmic density. Looking at Fig. 9, just by using this density cut, we can see that out of the initial ∼12 million haloes, we are left with only ∼1.5 million in this cosmic density range. This would mean that fewer than 12 per cent of all haloes reside in walls like our Local Wall. When we move to the next criterion in Fig. 9, 'radius cut + must have at least 1 halo within the stated mass range', this percentage plummets even further, with only 329 haloes in walls like out Local Wall within the cosmic density range considered. The percentage of haloes out of initial haloes at this cut is much less than 1 per cent. At this point, we could already conclude that our Local Wall is quite rare even without considering the other criteria that further reduce this percentage, albeit less dramatically.

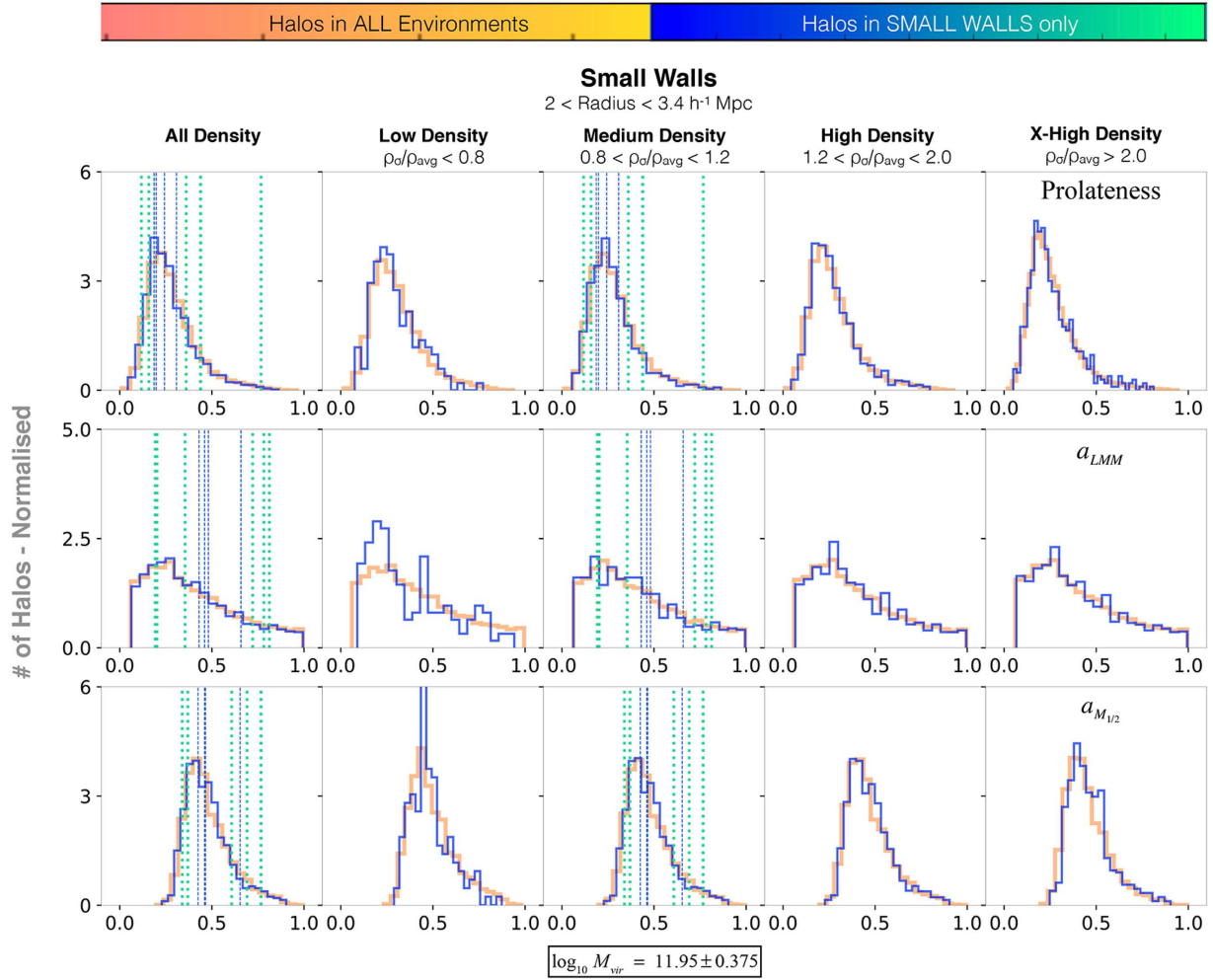

**Figure 11.** Histograms of halo properties in all web environments and small walls as a function of density smoothed on a scale of $8\,h^{-1}$ Mpc, for halo mass $\log_{10} M_{\rm vir}/M_\odot = 11.95 \pm 0.375$. *Left to right*: As in Fig. 10. *Top to bottom*: The distribution of prolateness $P$, scale factor of last major merger $a_{\rm LMM}$, and scale factor when the halo had half of its $z=0$ mass $a_{M_{1/2}}$ are compared with density in small walls (radius 2–3.4 $h^{-1}$ Mpc) (in blue) and environments of all sizes (orange). The histograms here corresponds to the second column of Fig. 3, although it is calculated at the smoothing scale of $8\,h^{-1}$ Mpc, and not $2\,h^{-1}$ Mpc of Fig. 3. Like the histograms of Fig. 10, the density range here of $0.8 < \rho_\sigma/\rho_{\rm avg} < 1.2$ corresponds to the average cosmic density (Klypin et al. 2015). We note that across all the different density bins, there is a general agreement in the distribution of halo properties between haloes in all environment, and haloes in walls. There appears to be a slight discrepancy for the low-density range of $\rho_\sigma/\rho_{\rm avg} < 0.8$, but this is due to the small number of haloes in walls at this range for good meaningful statistics: there are only 146 haloes in walls, while there are 13 631 haloes in all environment at the same range. For the other density ranges, there is a good agreement of properties across the different web environments. In the average cosmic density range of $0.8 < \rho_\sigma/\rho_{\rm avg} < 1.2$, we found 10 haloes that are most like those of the Milky Way Galaxy and Andromeda, which we illustrated with green dotted lines (near centre of wall) and blue dashed lines (otherwise). The halo properties here do not fall along a median as those in Fig. 10, as the properties here ($P$, $a_{\rm LMM}$ and $a_{M_{1/2}}$) are not known to follow any particular distribution, while those in the preceding histograms do ($\dot{M}/M$ follows a Gaussian distribution, while both $\lambda_{\rm B}$, and $C_{\rm NFW}$ follow a lognormal distribution). Here, we see that the properties of the 10 haloes most like our Milky Way galaxy and Andromeda are randomly placed, again indicating that these haloes properties of the Milky Way galaxy and Andromeda in our Local Wall are not peculiar.

For the 3 walls/6 haloes whose environments are most like that of our own Local Wall, we study their halo properties. In Figs 10 and 11, we looked at the distribution of halo properties in histograms of haloes of the mass $\log_{10} M_{\rm vir} = 11.95 \pm 0.375$ (∼mass of our Milky Way galaxy's dark matter halo), in small walls the size of the Local Group wall and calculated at the smoothing scale of $\rho_\sigma = 8\,h^{-1}$ Mpc where the average cosmic density is ∼1 (Klypin et al. 2015).

We drew six vertical dotted green lines to indicate where the halo properties of these 3 walls/6 haloes fall within the histogram. We also took a step backwards in the cut of haloes in Local Walls and added in the vertical blue dashed lines of 4 haloes to make up the 5 walls/10 haloes (that is, we do not restrict the haloes to just being in the centre of their walls) to provide more information on the distribution of the halo properties in environments like our Local Wall. We note that these halo properties fall around the median of each histogram with known distribution (Gaussian for the specific mass accretion rate, and approximately lognormal distributions (Rodríguez-Puebla et al. 2016) for $\lambda_{\rm B}$ and $C_{\rm NFW}$), while they fall randomly in histograms with no known distribution ($a_{\rm LMM}$, Prolateness, and $a_{M_{1/2}}$). Thus, the presence of haloes in walls like our Local Wall does not seem to affect the properties we studied. Hence, we conclude that although our own Local Group environment is somewhat special, it has no effect on the halo properties we have examined except for the orientation of halo angular momenta (Section 3.2.1).

## 4 SUMMARY AND CONCLUSION

We found in this paper that at a given environmental density, the different cosmic web environment of filaments, walls and voids does not have significant effects on any of the halo properties that we studied at $z = 0$: the halo mass accretion rate (dynamical time averaged) $\dot{M}_{\tau_{dyn}}/M$, spin parameter $\lambda_{\rm B}$, NFW concentration $C_{\rm NFW}$, prolateness $P$, scale factor of the last major merger $a_{\rm LMM}$, and scale factor when the halo had half of its $z = 0$ mass $a_{M_{1/2}}$. That is, the different locations of the cosmic web environment do not affect these core halo properties for haloes of the same mass and at the same environmental density, which is similar to the results of Romano-Diaz et al. (2017). We find that these halo properties are instead determined by the local environmental density of the halo.

In addition, we found that even though the presence of galaxies as massive as the Milky Way and Andromeda Galaxy near the centres of walls as small as our Local Wall is quite rare (0.03 per cent of all walls), it nevertheless appears to have essentially no effect on the halo properties that we studied. We also found that the angular momentum of haloes in walls tends to lie within about 30° of the walls, in agreement with Aragon-Calvo & Szalay (2013).

Our results in this paper are consistent with observational evidence of Yan et al. (2013), where the properties of galaxies at a given environmental density in the Sloan Digital Sky Survey do not depend on the cosmic web location, although as mentioned in Section 1 there are other authors who have found differences in galaxies in different cosmic web environments even at the same density, especially at higher redshifts.

It would be interesting to look at other halo properties in the Bolshoi–Planck or other simulations, to see if the web environment has any effects on them at constant density. In addition, it would also be interesting to look at halo properties at earlier time-steps $z > 0$. Lastly, it would also be interesting to examine galaxies in hydrodynamic simulations to see whether the galaxies have cosmic web dependences at fixed environmental density, unlike our results regarding dark matter haloes.

## ACKNOWLEDGEMENTS

We thank Avishai Dekel, Sandra Faber, Duncan Farrah, Marshall McCall, Rachel Somerville, Paul Sutter, and Simon White for helpful discussions, including at the 2017 Galaxy-Halo workshop at KITP. We also thank David Spergel, Jacqueline van Gorkom, and Mordecai Mac-Low for stimulating discussions comparing our theoretical conclusion with observational results, and Shy Genel for follow-up steps to a new project. Lastly, we also thank the Galaxies Group and Stream Team of Columbia University for providing numerous critiques. Computational resources supporting this work were provided by the NASA High-End Computing (HEC) Program through the NASA Advanced Supercomputing (NAS) Division at Ames Research Centre, and by the Hyades astrocomputer system at UCSC. We also benefitted from use of the 3D Astro-Visualisation Lab at UCSC. JP acknowledges support from grant HST-AR-14578.001-A. JP and ARP acknowledge the UC MEXUS-CONACYT Collaborative Research Grant CN-17-125.

## REFERENCES

Alam S., Zu Y., Peacock J., Mandelbaum R., 2018, MNRASpre-print (arXiv:1801.04878)
Allgood B., Flores R. A., Primack J. R., Kravtsov A. V., Wechsler R. H., Faltenbacher A., Bullock J. S., 2006, MNRAS, 367, 1781
Alonso D., Eardley E., Peacock J. A., 2015, MNRAS, 447, 2683
Alonso D., Hadzhiyska B., Strauss M., 2016, MNRAS, 460, 256
Aragon-Calvo M. A., Szalay A. S., 2013, MNRAS, 428, 3409
Aragon-Calvo M. A., Yang L. F., 2014, MNRAS, 440, L46
Aragón-Calvo M. A., Platen E., van de Weygaert R., Szalay A. S., 2010, ApJ, 723, 364
Behroozi P. S., Weschler R. H., Wu H.-Y., 2013a, ApJ, 762, 109
Behroozi P. S., Weschler R. H., Wu H.-Y., Klypin A., Primack J., 2013b, ApJ, 763, 15
Blumenthal G. R., Faber S. M., Primack J. R., Rees M. J., 1984, Nature, 311, 517
Bond J. R., Kofman L., Pogosyan D., 1996, Nature, 380, 603
Brouwer M. M. et al., 2016, MNRAS, 462, 4451
Bullock J. S., Dekel A., Kolatt T. S., Kravtsov A. V., Klypin A. A., Porciani C., Primack J. R., 2001, ApJ, 555, 240
Doroshkevich A. G., Shandarin S. F., Zeldovich I. B., 1983, in Abell G. O., Chincarini G., eds, IAU Symp. Vol. 104, Early Evolution of the Universe and its Present Structure, Cambridge Univerity Press, p. 387
Eardley E. et al., 2015, MNRAS, 448, 3665
Guo Q., Tempel E., Libeskind N. I., 2015, ApJ, 800, 112
Hahn O., Porciani C., Carollo C. M., Dekel A., 2007, MNRAS, 375, 489
Klypin A., Karachentsev I., Makarov D., Nasonova O., 2015, MNRAS, 454, 1798
Klypin A., Yepes G., Gottlöber S., Prada F., Heß S., 2016, MNRAS, 457, 4340
Laigle C. et al., 2015, MNRAS, 446, 2744
Laigle C. et al., 2018, MNRAS, 474, 5437
Lee C. T., Primack J. R., Behroozi P., Rodríguez-Puebla A., Hellinger D., Dekel A., 2017a, MNRAS, 466, 3834
Lee C. T., Primack J. R., Behroozi P., Rodríguez-Puebla A., Hellinger D., Dekel A., 2017b, preprint (arXiv:e-prints)
Libeskind N. I., Hoffman Y., Steinmetz M., Gottlöber S., Knebe A., Hess S., 2013, ApJ, 766, L15
Libeskind N. I. et al., 2018, MNRAS, 473
McCall M. L., 2014, MNRAS, 440, 26
Navarro J. F., Frenk C. S., White S. D. M., 1996, ApJ, 462, 563
Navarro J. F., Abadi M. G., Steinmetz M. M., 2004, ApJ, 613, L41
Neyrinck M., 2008, MNRAS, 386, 10
Peebles P. J. E., 1969, ApJ, 155, 393
Planck Collaboration XXVI, 2014, A&A, 571, 66
Planck Collaboration XIII, 2016, A&A, 594, A13

Platen E., van de Weygaert R., Jones B. J. T., 2007, MNRAS, 380, 551
Rodríguez-Puebla A., Behroozi P., Primack J., Klypin A., Lee C., Hellinger D., 2016, MNRAS, 462, 893
Romano-Diaz E., Borzyszkowski M., Porciani C., E. G., 2017, MNRAS, 469, 1809
Sutter P. M. et al., 2015, Astron. Comput., 9, 1
Tojeiro R., 2017, MNRAS, 470, 3720
van de Weygaert R., Shandarin S., Saar E., Einasto J., 2016, IAU Symp. Vol. 308, The Zeldovich Universe: Genesis and Growth of the Cosmic Web, Cambridge Univerity Press
Wang P., Kang X., 2017, MNRAS, 468, L123
Yan H., Fan Z., White S. D. M., 2013, MNRAS, 430, 3432
Zel'dovich Y. B., 1970, A&A, 5, 84

## APPENDIX A: WALLS OF ALL SIZES

This appendix contains figures that supplement those in the text. Like Figs 2 and 3, Figs A1 and A2 show the median halo properties in different cosmic environments, but now in walls of different sizes, ranging from small ($2$–$3.4\,h^{-1}$ Mpc) to extremely large ($> 9.5\,h^{-1}$ Mpc). We see that the halo properties are similar across walls of various sizes, indicating that the size of the walls does not affect those properties. They appear to be only affected by the local environmental density. Similarly, like Figs 4 and 5, Figs A3 and A4 show the full distribution of halo properties in different cosmic environments, but again in walls of different sizes. Again, we do not see a difference in distribution of halo properties due to the sizes of the walls.

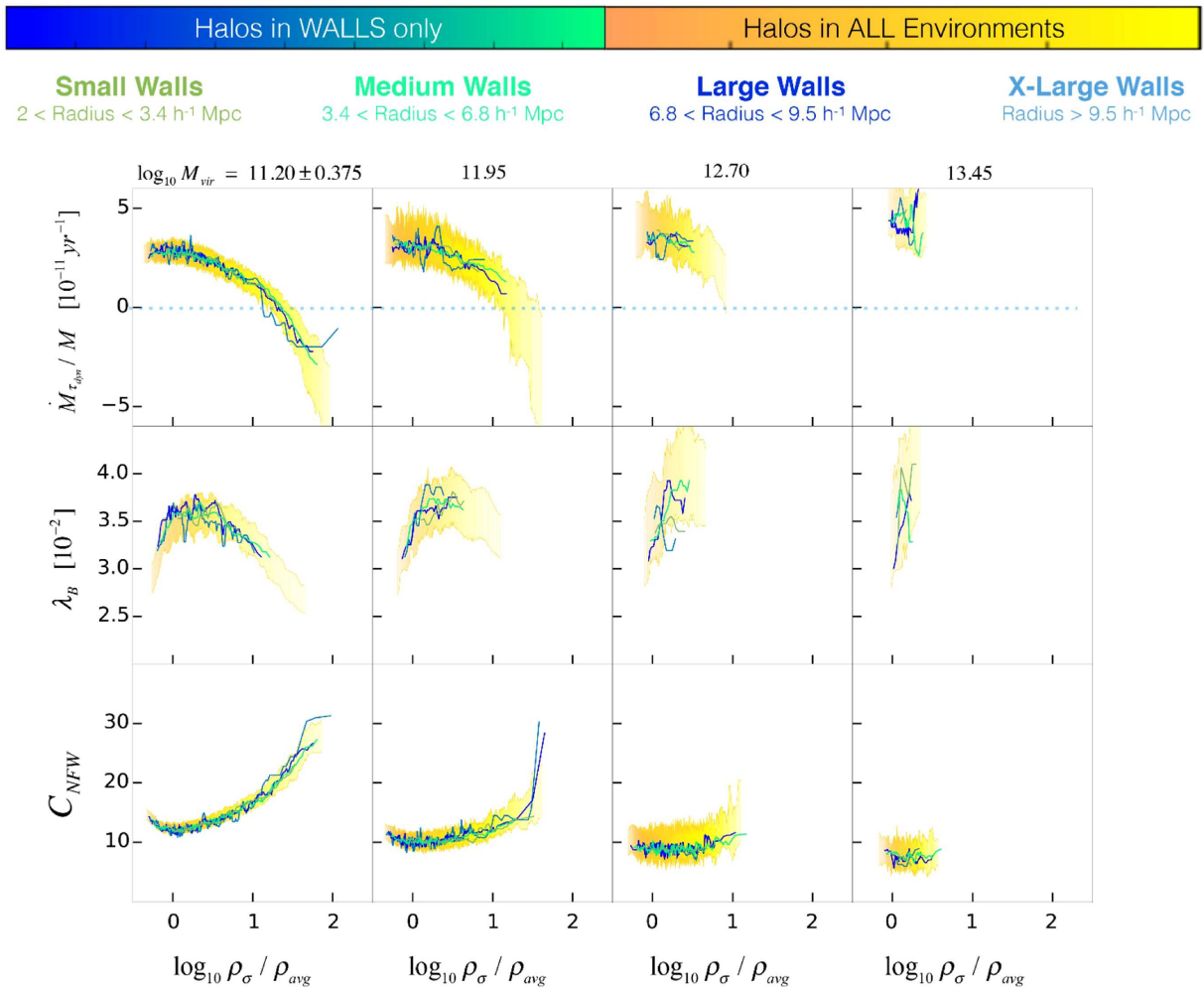

**Figure A1.** Halo properties in all-sized walls as a function of density. *Left to right*: The columns represent four mass bins with corresponding smoothing scales, as in Fig. 2. *Top to bottom*: The median distribution of specific halo mass accretion rate $\dot{M}/M$, spin parameter $\lambda_B$, and concentration $C_{NFW}$ are plotted versus density in different mass bins across different-sized walls. The scatter plots in green to blue shading are the median halo properties corresponding to walls of different sizes. We split the walls up into different sizes because we are most interested in small ($2$–$3.4\,h^{-1}$ Mpc) walls, as that corresponds to the size of our Local Wall (McCall 2014). These scatter plots fall within the thick yellow band, which represents the 5th–95th percentile dispersion of the median of each halo property in all web environments. The right tail-end of each sub-plot has the fewest haloes, and any differences at these tail-ends between the different-sized walls are not statistically significant. *Results*: We note that only the first two mass bins/columns have sufficient median haloes in all walls for meaningful results. We see that although there are tiny deviations, the halo properties do not appear to be affected on the whole by different-sized walls, but are affected by increasing density. This is analogous to the results of Fig. 2, where we looked at halo properties across different web environments including small walls. Similar to Fig. 2, the horizontal blue dotted line in the specific accretion rate plots separates the haloes gaining and losing mass.

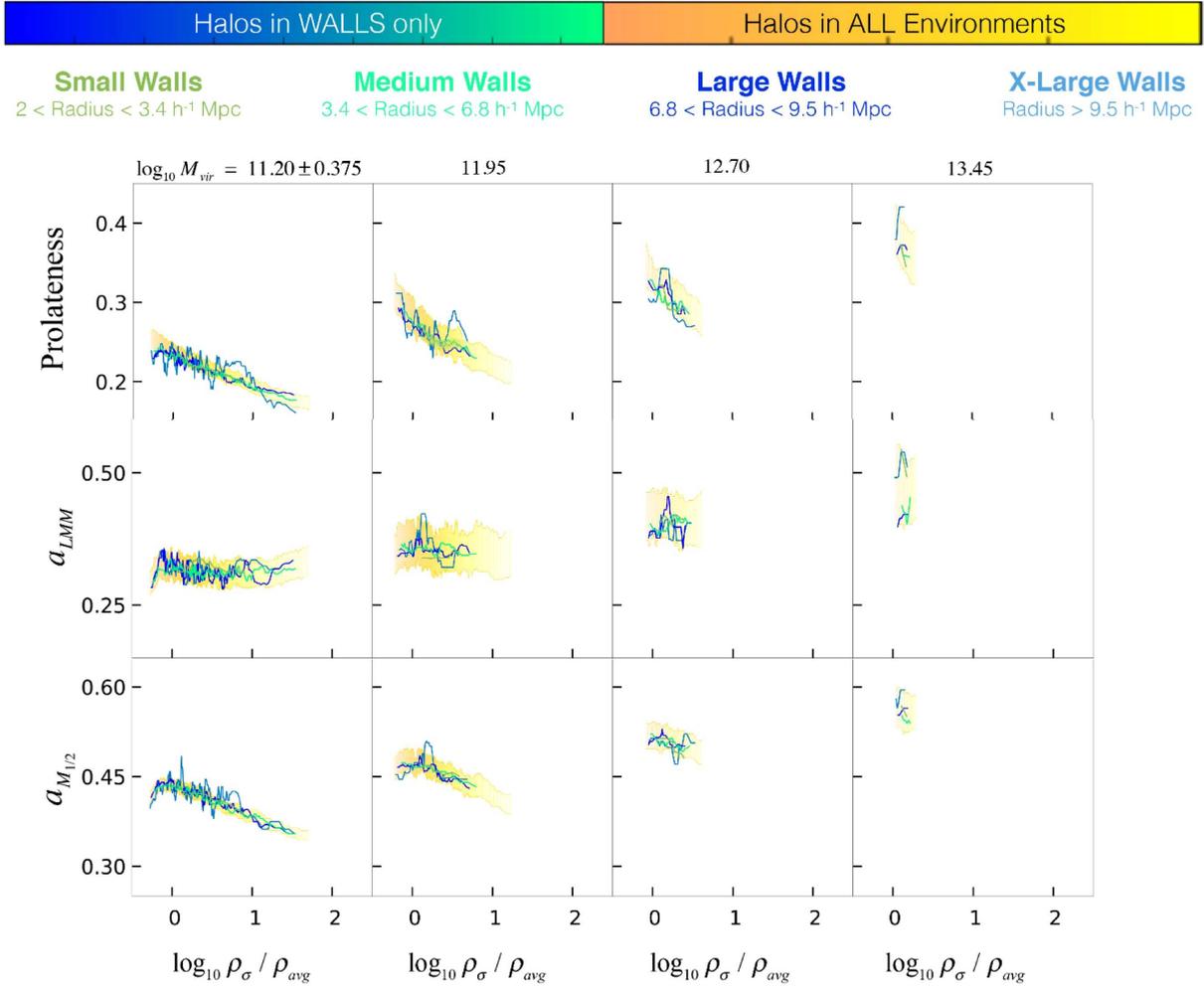

**Figure A2.** Halo properties in all-sized walls as a function of density. *Left to right*: The columns represent four mass bins with corresponding smoothing scales, as in Fig. 2. *Top to bottom*: The median distribution of prolateness, scale factor of last major merger, and scale factor when the halo had half of its $z = 0$ mass are plotted versus density in different mass bins across different-sized walls. *Results*: We note that only the first two mass bins/columns have sufficient median haloes in all walls for meaningful results. Similar to the results of Fig. A1, we see that the halo properties does not appear to be affected on the whole by the size of the walls. In addition, the scale factor of the last major merger does not seem to be affected by increasing density, although prolateness and the half-mass scale do. This is analogous to the results of Fig. 3, where we looked at halo properties across different web environments including small walls.

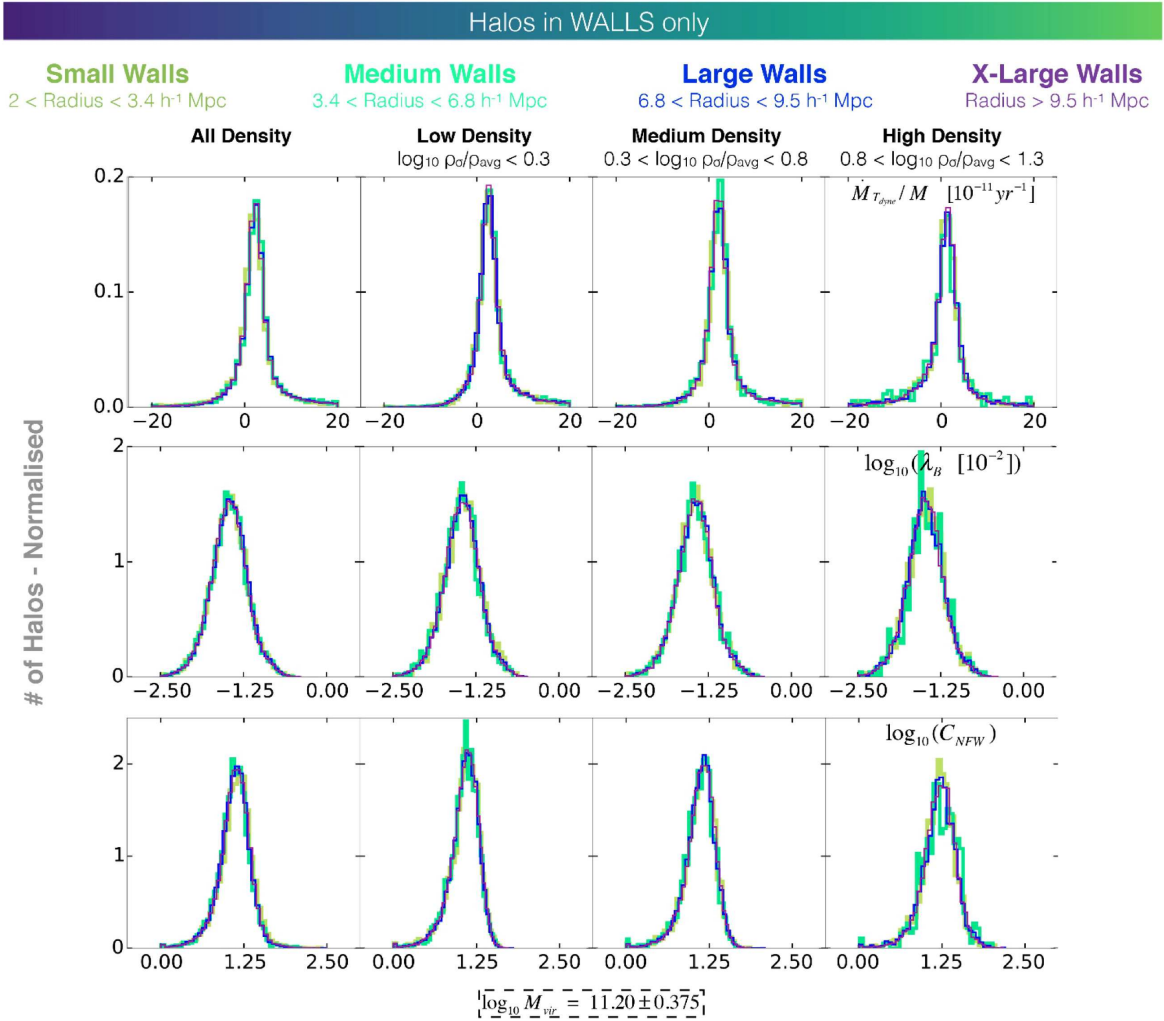

**Figure A3.** Halo properties in walls of all sizes as a function of environmental density, for halo mass $\log_{10} M_{\rm vir}/M_\odot = 11.20 \pm 0.375$. *Left to right*: We split density into three different regions to explore the effect of low to high density on the halo properties. *Top to bottom*: Specific accretion rate, $\lambda_{\rm B}$, and $C_{\rm NFW}$ are plotted versus density in walls of all sizes. The histograms here correspond to those of Fig. 4 of this paper, where we showed histograms of properties in different cosmic environments and small walls instead. *Results general*: The halo properties do not appear to be affected by the different sizes of the walls. We note that that halo properties in the small and medium walls here appear to deviate from the distribution, taking on a more 'jaggedy' appearance, as we go towards larger densities due to lack of haloes at these larger densities. This means that there is a lack of haloes in small and medium walls at high densities, with the lack of haloes contributing to the deviation. However, despite this deviation, we can see that the trend of the halo properties in small and medium walls generally follows that in walls of all sizes for each property.

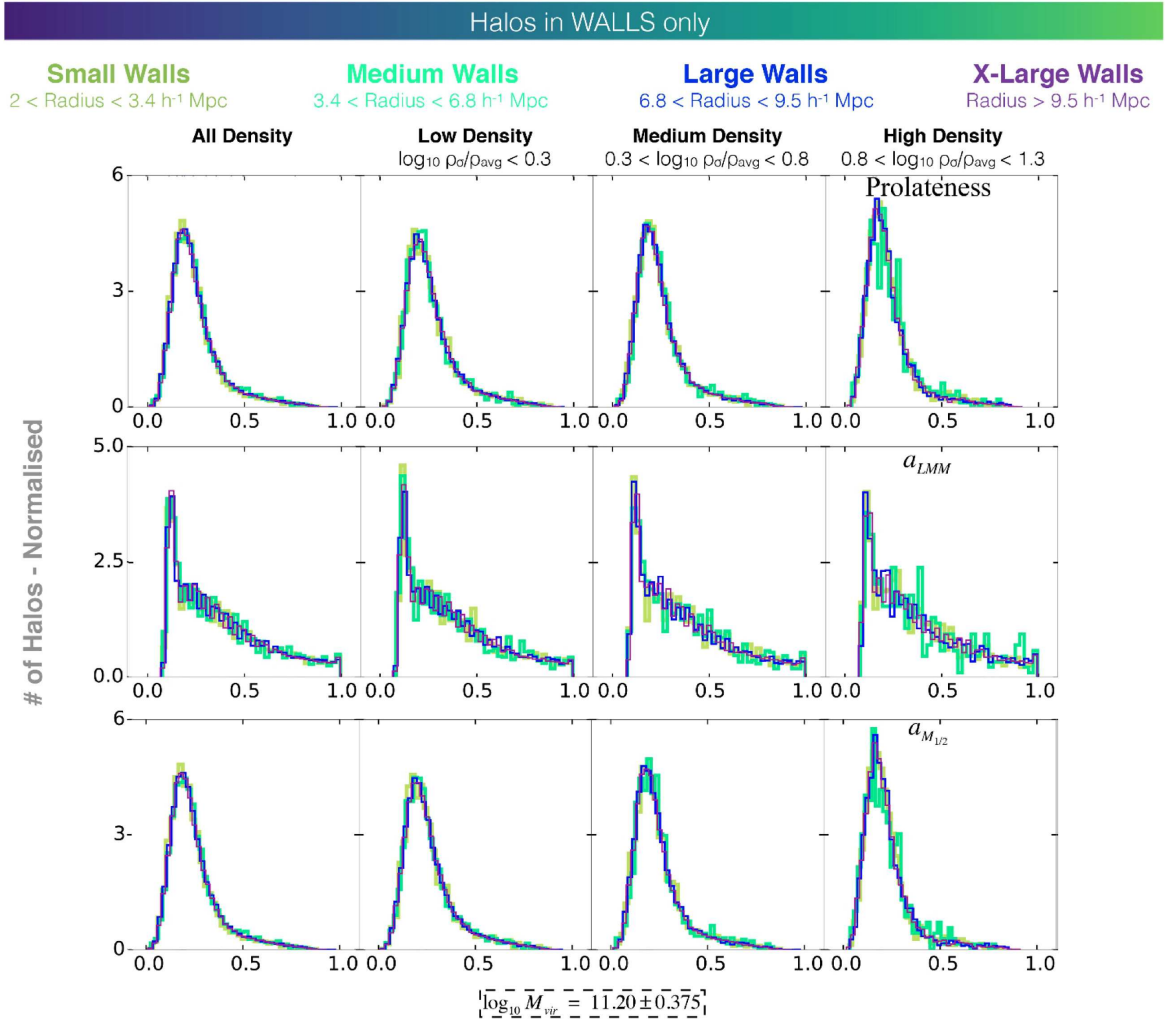

**Figure A4.** Histograms of halo properties in all-sized walls as a function of density, for halo mass $\log_{10} M_{\rm vir}/{\rm M}_\odot = 11.20 \pm 0.375$. *Left to right*: As in Fig. A3. *Top to bottom*: The full distribution of prolateness, scale factor of last major merger, and the scale factor when the halo had half of its $z = 0$ mass are plotted in walls of all sizes. The histograms here correspond to those of Fig. 5 of this paper, where we showed histograms of properties in different cosmic environments and small walls instead. *Results*: The halo properties do not appear to be affected by the different sizes of the walls. Similar to Fig. A3, the small and medium walls here appear to deviate from the distribution as we go towards higher densities due to lack of haloes at these high densities. However, despite this, we can see the trend of the haloes in small and medium walls generally following the haloes in walls of all sizes for each property. Hence, along with the histograms of Fig. A3, we conclude that wall sizes do not significantly affect their halo properties.

## APPENDIX B: CHANGING PARAMETER D

For all of the distributions in the paper, we have used a parameter $D = 0.25\,h^{-1}$ Mpc as the distance away from a filament, together with $< 0.75\,h^{-1}$ Mpc as the distance to a wall, to group haloes into the environment we called a 'wall'; those haloes not in filaments or walls are considered to be in voids. In this appendix, we show the effect when we change $D$ to be $D = 0.75\,h^{-1}$ Mpc instead. We can see that the effects are quite small, strengthening our argument that the cosmic environment has little effect on halo properties; it is density which governs the halo properties. It should be noted that we did not use $< 0.75\,h^{-1}$ Mpc as the distance to a wall when looking for haloes like those in our Local Wall. Not using this criterion has the effect of limiting haloes to just the plane of the walls, which is the case for the Milky Way Galaxy and Andromeda (McCall 2014).

### Curve-fitting to specific accretion rate for all densities

| Accretion Rate [$10^{-11}$ yr$^{-1}$] | Median | Mean | Standard Deviation | Model |
|---|---|---|---|---|
| Small Walls | 1.765 | 2.201 | 2.251 | Gaussian |
| Medium Walls | 2.678 | 2.382 | 2.228 | Gaussian |
| Large Walls | 2.530 | 2.338 | 2.280 | Gaussian |
| X-Large Walls | 2.605 | 2.230 | 2.209 | Gaussian |
| All | 2.500 | 2.093 | 2.400 | Gaussian |
| Voids | 2.497 | 2.412 | 2.322 | Gaussian |

### Curve-fitting to $\log_{10} \lambda_B$ for all densities

| $\log_{10} \lambda_B$ [$10^{-2}$] | Median | Mean | Standard Deviation | Model |
|---|---|---|---|---|
| Small Walls | -1.433 | -1.451 | 0.260 | Lognormal |
| Medium Walls | -1.338 | -1.462 | 0.254 | Lognormal |
| Large Walls | -1.473 | -1.451 | 0.260 | Lognormal |
| X-Large Walls | -1.470 | -1.458 | 0.261 | Lognormal |
| All | -1.457 | -1.469 | 0.264 | Lognormal |
| Voids | -1.460 | -1.506 | 0.268 | Lognormal |

### Curve-fitting to $\log_{10} C_{NFW}$ for all densities

| $\log_{10} C_{NFW}$ | Median | Mean | Standard Deviation | Model |
|---|---|---|---|---|
| Small Walls | 1.157 | 1.145 | 0.200 | Lognormal |
| Medium Walls | 1.086 | 1.111 | 0.193 | Lognormal |
| Large Walls | 1.097 | 1.123 | 0.198 | Lognormal |
| X-Large Walls | 1.155 | 1.131 | 0.197 | Lognormal |
| All | 1.154 | 1.147 | 0.206 | Lognormal |
| Voids | 1.146 | 1.142 | 0.181 | Lognormal |

**Figure B1.** Using parameter $D = 0.75 \, h^{-1}$ Mpc away from a filament, these are curve-fittings of the distributions of specific accretion rate $\dot{M}/M$, spin parameter $\lambda_B$, and concentration $C_{NFW}$ with haloes $> 0.75 \, h^{-1}$ Mpc away from filaments, and $< 0.75 \, h^{-1}$ Mpc away from walls. *Results*: With a larger distance away from filaments, the haloes represented in this table are more concentrated in the centre of the walls than those in Fig. 6. We note that the numbers tabulated here are very close to those found in Fig. 6, indicating that the distribution of histograms in Figs 4 and 5 and the median plots in Figs 2 and 3 are very similar whether using the halo-to-filament distance parameter $D = 0.25 \, h^{-1}$ Mpc or $D = 0.75 \, h^{-1}$ Mpc. We used a table of fits to show the similarity between the two different parameters, as the plots and histograms using these two parameters are too similar to see by eye the differences between them. We have, however, still included the median plots just for comparison in B2.

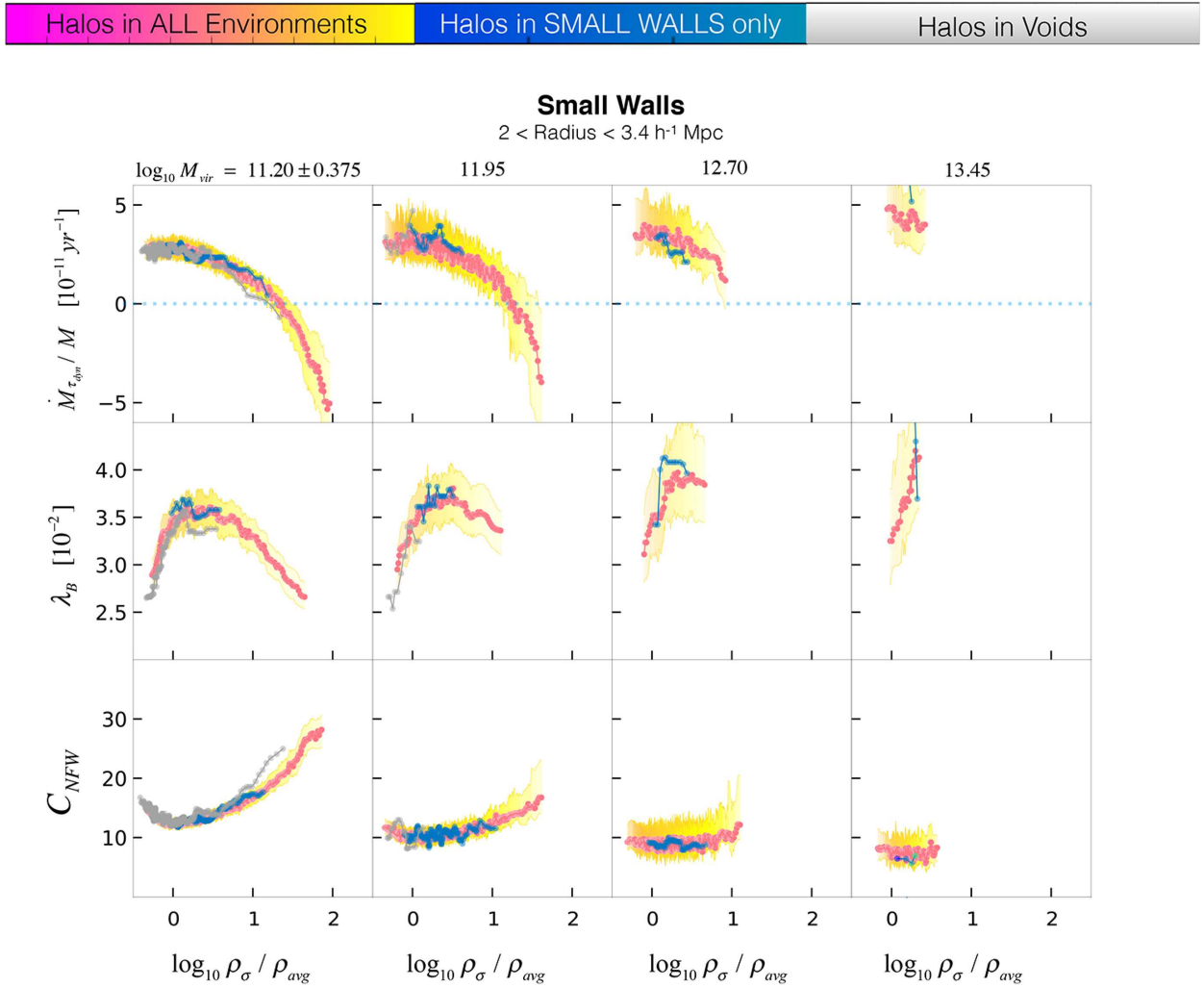

**Figure B2.** Using parameter $D = 0.75\,h^{-1}$ Mpc away from a filament, these are halo properties in all web environments, small walls, and voids as a function of density, where $\rho_\sigma$ is density used on that smoothing scale. *Left to right*: The columns have been split into four mass bins of $\log_{10} M_{vir}/M_\odot = 11.20 \pm 0.375$, $11.95 \pm 0.375$, $12.70 \pm 0.375$, and $13.45 \pm 0.375$, with density smoothing scales of $\rho_\sigma$ in these mass bins as 1, 2, 4, and 8 $h^{-1}$ Mpc, respectively. *Top to bottom*: The median distribution of specific accretion rate, $\lambda_B$, and $C_{\rm NFW}$ are plotted versus density in the four mass bins. *Results*: The plots here uses the parameter $D = 0.75\,h^{-1}$ Mpc, while those in Fig. 2 uses $D = 0.25\,h^{-1}$ Mpc. Looking between these two plots, it is difficult to see by eye the differences arising from using the different halo-distance-to-filaments parameter $D$. It is better to make the comparison using the quantities found in the table in Fig. B1 instead.

# APPENDIX C: ANGULAR MOMENTUM ORIENTATION OF HALOES IN THE CENTRES OF WALLS

In Section 3.2.1, we used the following cuts to constraint the number of haloes: walls of the Bolshoi–Planck:

(i) Radius (wall): 2–3.4 $h^{-1}$ Mpc
(ii) Mass (haloes): 0.8–1.2 × $10^{12}$ M$_\odot$
(iii) Distance of halo to filaments: >0.25 $h^{-1}$ Mpc

Here, we show an alternative criterion for the haloes, by limiting them to the centre of their walls. We define here the haloes to be near the centre of the wall with the following criteria:

(i) radius (wall): 2–3.4 $h^{-1}$ Mpc,
(ii) mass (haloes): 0.8–1.2 × $10^{12}$ M$_\odot$,
(iii) distance of halo to centre of mass of the wall: <1.75 $h^{-1}$ Mpc,

where we calculated the centre of mass as the mean of the position voxels of each wall. These criteria limit the number of haloes to 123 in 110 walls. This criteria of limiting haloes to the centre of walls here produces fewer haloes than those found using the methods in Section 3.2.1, where we found 702 haloes in 594 walls by reassigning haloes in walls, but not limiting them to be near the centres of their walls. The results found here of the dot product between the orthogonal vector and the $\omega$ vector is 0.458, while those in Section 3.2.1 is 0.452. Hence, though we have limited the angular momentum to haloes near the centres of walls, the qualitative result for the direction of the angular momentum vector lying close to the plane of the wall remains the same as that of for haloes not stringently confined to the centres of their walls.

This paper has been typeset from a TEX/LATEX file prepared by the author.